\documentclass[11pt,a4paper,english]{article}
\usepackage{float}
\usepackage{graphicx}
\usepackage{amsmath}

\usepackage{amsfonts}
\usepackage{latexsym}

\usepackage{babel}

\newcommand{\beq}{\begin{equation}}
\newcommand{\eeq}{\end{equation}}

\newcommand{\intd}{\mathrm{d}}

\newcommand{\mc}[1]{\mathcal{#1}}

\newcommand{\CC}{\mathbb{C}}
\newcommand{\ZZ}{\mathbb{Z}}

\newcommand{\ii}{\mathrm{i}}

\newcommand{\ket}[1]{|#1 \rangle}
\newcommand{\braket}[2]{\langle #1 | #2 \rangle}

\newcommand{\brakettt}[3]{\langle #1 | #2 |#3 \rangle}
\newcommand{\braketv}[2]{\langle #1 , #2 \rangle}

\newcommand{\mass}{\mu}

\makeatletter \@addtoreset{equation}{section} \makeatother
\renewcommand{\theequation}{\arabic{section}.\arabic{equation}}

\begin{document}

\title{{\bf Higher spin fields with reversed\\ spin-statistics relation}}

\author{
{}\\[3mm]
G\'abor Zsolt T\'oth
\\[6mm] 
\small \it Institute for Particle and Nuclear Physics, Wigner Research Centre for Physics, \\
\small \it MTA Lend\"ulet Holographic QFT Group, Konkoly Thege Mikl\'os \'ut 29-33,\\
\small \it 1121 Budapest, Hungary\\[1mm]
\small  E-mail: \tt toth.gabor.zsolt@wigner.mta.hu
\date{}
}
\maketitle

\vspace{1cm}
\begin{abstract}
A construction of massive free fields with arbitrary spin 
and reversed spin-statistics relation is presented.
The main idea of the construction is to consider fields that transform according to  
representations of the Lorentz group that are doubled in comparison with 
the representations according to which 
normal (physical) fields transform.
This allows the definition of opposite 
commutation properties for these fields, 
while the spin of the particles they describe remains unchanged. 
The correspondence established by the construction between fields obeying 
normal and reversed spin-statistics relation makes it possible to express 
e.g.\ 
the polarization states, (anti)commutators, or Feynman propagators of the 
latter fields in terms of those of the normal fields to which they correspond. 
The cases of the scalar and Dirac fields are discussed in additional detail.
\end{abstract}

\vspace{1cm}
{\small
\hspace{2.3mm}
Journal reference: International Journal of Modern Physics A 29 (2014) 1450129
}

\thispagestyle{empty}

\newpage

\section{Introduction}
\label{sec.introd}

It is well-known that in quantum field theory, in addition to clearly physical fields, 
other fields that are not in every respect physical also have important role. 
For example, the Faddeev--Popov ghost fields \cite{FP, WeinbergQFT2, PeskinSchroeder} appearing in
quantized gauge field theories are of this kind.
The main non-physical features of these fields are that although they are scalar fields,
they are fermionic (anticommuting), and the Fock space generated by their modes has indefinite scalar product. 
Non-zero spin ghost fields with similar features also appear 
in certain theories that describe higher spin fields, for example
vector and spinor ghost fields appear 
in quantum gravity and supergravity \cite{DeWitt, Superspace}.
Ghost fields are massless in gauge theories with unbroken gauge symmetry, but they
can acquire mass due to spontaneous symmetry breaking \cite{WeinbergQFT2, PeskinSchroeder}, and massive  
ghost fields appear also in some other massive extensions of gauge theories \cite{CF, Kondo, Kondo2}.

In this paper an elementary construction of massive free fields with arbitrary spin and reversed spin-statistics relation is presented. 
In relation to the latter property, the Fock space generated by the modes of these fields also has indefinite scalar product.
Regarding other properties, specifically locality and the non-negativity of the energy spectrum,
the fields with reversed statistical properties are not different from normal fields.

The fields with reversed spin-statistics properties 
are constructed using data that specifies normal fields.
This shows, in particular, that to any normal field a corresponding field with the same spin but
reversed spin-statistics relation
can be associated. The main idea of the construction is 
to double the Lorentz group representations according to which the fields transform, 
whereby it becomes possible to prescribe fermionic commutation properties for integer spin fields 
and bosonic commutation properties for half-integer spin fields.

The construction is given in a formalism that was introduced in Ref. \cite{toth}, 
in which elementary (irreducible) higher spin fields $\phi_{(s)}$ are obtained in the form 
\beq
\label{eq.bev}
\phi_{(s)\alpha} = {\mc{D}[\partial]_\alpha}^\beta \Phi_\beta\ ,
\eeq
where $\Phi$ is an auxiliary multi-component Klein--Gordon field, which satisfies canonical (anti)-commutation relations,
and $\mc{D}[\partial]$ is a differential operator that projects out the appropriate degrees of freedom from $\Phi$.
This formalism provides canonical field variables and thus it allows the 
derivation of canonical Hamiltonian equations of motion, if the interaction Hamiltonian operator as a 
local expression of fields is known. In this respect the formalism of Ref. \cite{toth} extends that of Refs. \cite{Weinberg,WeinbergQFT}. 
The calculation of Feynman propagators, (anti)commutators and Green functions of higher spin fields is also simple
if they are given in the form (\ref{eq.bev}). A further feature of the formalism of Ref. \cite{toth} is that it exhibits 
the reality properties of the fields and of various tensors in a clear and covariant manner.

We focus, as in Ref. \cite{toth}, on normal fields transforming according to 
the representations $(n/2,m/2)\oplus (m/2,n/2)$ and $(n/2,n/2)$, $n,m\in\ZZ$, $n,m\ge 0$,
$n\ne m$,
of $SL(2,\CC)$ (the covering group of the Lorentz group), 
since these are the irreducible real representations of this group, 
and apply the construction to these fields. Nevertheless, 
the construction can be applied to 
other types of fields, 
like the Rarita--Schwinger \cite{RS} or the Wigner--Bargmann \cite{BW}
fields, in a similar way.

Our aim is mainly to obtain an overview of the fields that may in principle appear in a quantum field theoretical model,
and specifically to extend the survey of fields given in Refs. \cite{Weinberg,Weinberg3} 
(also presented in the textbook \cite{WeinbergQFT}) by considering 
fields with reversed spin-statistics relation.
We stress that we do not make any restrictive assumption about the role of these fields; 
in particular we do not assume that they are Faddeev--Popov--DeWitt type ghost fields. 

Higher spin fields have attracted much interest for several decades, and their investigation continues also
in the present days. For an incomplete collection of recent results 
we refer the reader to Refs. \cite{JPhysA} - \cite{NG}. 
The constant interest in higher spin fields can be explained partially by the fact that 
it is not easy to construct consistent models for their interactions.  
We restrict ourselves to the construction of free fields and thus deal with a much simpler problem, nevertheless 
we hope that our elementary discussion can still be of interest. 
As far as we know, a similar discussion  
of higher spin fields with reversed spin-statistics relation 
cannot be found in the literature.

The paper is organized as follows.
In Section \ref{sec.1} we recall in a shortened form the framework that we presented in Ref. \cite{toth} for describing normal fields. 
In Section \ref{sec.RSS} we present our construction of higher spin fields with reversed spin-statistics relation.    
In Sections \ref{sec.scRSS} and \ref{sec.dirac} we discuss the special cases of the scalar and Dirac fields, respectively. 
A summary is given in Section \ref{sec.conclusion}.
Appendix \ref{sec.feynman} contains formulas for the Feynman propagators, Green functions and (anti)commutation relations of 
the various fields.
In Appendix \ref{sec.appb} certain technical details about polarization vectors are given.

The signature of the Minkowski metric $g_{\mu\nu}$ is taken to be $({+}{-}{-}{-})$. 
Upper indices are used for Minkowski (space-time)
vectors and lower indices for Minkowski covectors.
Spinor indices and similar indices that label vector components in the representation spaces of $SL(2,\CC)$ are
often written explicitly, but they are suppressed at other places in the text, depending on which notation appears clearer.  
Such indices are not raised or lowered. For vector spaces and representations the star ${}^*$ is used to denote the 
dual vector space and representation.    
The dagger ${}^\dagger$ is used to denote the adjoint of operators in Hilbert spaces, and an analogous operation in the context of Lagrange formalism. 
The term Poincar\'e group refers in this paper
to the simply connected covering group of the usual connected Poincar\'e group generated by the translations and Lorentz transformations
of the Minkowski space-time.

\section{Formalism for higher spin fields}
\label{sec.1}

\subsection{Multi-component Klein--Gordon fields}
\label{sec.mkg}

The first step of the construction of higher spin fields described in Ref. \cite{toth} is to introduce 
multi-component (vector valued) bosonic and fermionic (i.e.\ commuting and anticommuting) real Klein--Gordon fields
$\Phi_\alpha$, which we will also refer to as auxiliary fields. These fields are the fundamental 
objects from which other kinds of fields are derived.  
They are also to some extent analogous to the 
electromagnetic vector potential in the Gupta--Bleuler formalism.

The Hamiltonian operator for the multi-component real Klein--Gordon fields is 
\begin{eqnarray}
\label{eq.H}
H & = & \frac{1}{2} \int \intd^3 x\, :\big[ \Pi_\alpha \Pi_\beta \epsilon^{\alpha\beta}  + (\partial_x\Phi_\alpha) (\partial_x\Phi_\beta) \epsilon^{\alpha\beta}  + \mass^2\Phi_\alpha\Phi_\beta \epsilon^{\alpha\beta} \big]:\ ,
\end{eqnarray}
where $\mass$ is the mass parameter and $\Pi_\alpha$ is the momentum field corresponding to $\Phi_\alpha$. 
The components $\Phi_\alpha$ of $\Phi$ are indexed by $\alpha=1,\dots,N$. 
In the fermionic case $N$ is even and $\epsilon^{\alpha\beta}$ 
is a nondegenerate antisymmetric and purely imaginary matrix: 
\begin{eqnarray}
\label{eq.eps1}
\epsilon^{\alpha\beta} & = & -\epsilon^{\beta\alpha}\\
\label{eq.eps2}
\epsilon^{\alpha\beta} & = & -\epsilon^{\alpha\beta *}\ .
\end{eqnarray}
In the bosonic case $\epsilon^{\alpha\beta}$ is a nondegenerate symmetric and real matrix: 
\begin{eqnarray}
\label{eq.eps3}
\epsilon^{\alpha\beta} & = & \epsilon^{\beta\alpha}\\
\label{eq.eps4}
\epsilon^{\alpha\beta} & = & \epsilon^{\alpha\beta *}\ .
\end{eqnarray}
We denote the inverse of $\epsilon^{\alpha\beta}$ by $\epsilon_{\alpha\beta}$; thus 
\beq
\epsilon_{\alpha\beta}\epsilon^{\beta\gamma}={\delta_\alpha}^\gamma\ .
\eeq
$\epsilon^{\alpha\beta}$ and $\epsilon_{\alpha\beta}$ are not used to raise or lower indices.  
$\Phi_\alpha$ and $\Pi_\alpha$ satisfy, by definition, the canonical equal time (anti)commutation relations
\beq
\label{eq.ac1}
[ \Phi_\alpha(x,t),\Pi_\beta(x',t) ]_\pm  =  \ii \epsilon_{\alpha\beta}\delta^3(x-x')\ ; \\
\eeq
all other (anti)commutators of $\Phi_\alpha$ and $\Pi_\alpha$ are $0$.
Here and in the following the notation $[\ ,\ ]_\pm$ is used to indicate that either 
a commutator (corresponding to the sign $-$) or an anticommutator (corresponding to the sign $+$) is meant;
commutators apply to the case of bosons
and anticommutators to the case of fermions.

Due to the commutation properties of $\Phi_\alpha$, in the bosonic case 
the antisymmetric part of $\epsilon^{\alpha\beta}$, if it had one, would not enter into (\ref{eq.H}), 
and the same applies to the symmetric part of $\epsilon^{\alpha\beta}$ in the fermionic case. Therefore requiring
the symmetry properties (\ref{eq.eps1}) and (\ref{eq.eps3}) does not cause any loss of generality. 
If $\epsilon^{\alpha\beta}$ has the symmetry property (\ref{eq.eps1}) or (\ref{eq.eps3}), then 
the reality property (\ref{eq.eps2}) or (\ref{eq.eps4}) follows from the requirement that 
the Hamiltonian operator (\ref{eq.H}) should be self-adjoint.

$\Phi_\alpha$ has the mode expansion 
\beq
\Phi_\alpha (x,t)=\int \frac{\intd^3 k}{\sqrt{2}(\sqrt{2\pi})^3 \omega(k)} \big[e^{\ii kx}e^{\ii \omega(k)t}
      a_\alpha^\dagger(k)
+e^{-\ii kx}e^{-\ii\omega(k)t}
      a_\alpha (k)\big]\ ,
\eeq
where 
$a_\alpha^\dagger (k)$ and $a_\alpha(k)$ are 
creation and annihilation operators, respectively, $k$ denotes $3$-dimensi-onal momentum, and $\omega(k)=\sqrt{\mass^2+k^2}$. 
The non-zero (anti)commutators of these creation and annihilation operators are
\beq
\label{eq.cc1}
[ a_\alpha(k),a_\beta^\dagger (k') ]_\pm  =  \epsilon_{\alpha\beta}\delta^3(k-k')\omega(k)\ .
\eeq
The Hamiltonian operator can be expressed in terms of these operators as
\begin{eqnarray}
\label{eq.H2}
H & = & \int \frac{\intd^3 k}{\omega(k)}\ \omega(k)\, a_\alpha^\dagger(k) a_\beta(k) \epsilon^{\alpha\beta}\ .
\end{eqnarray}
The Heisenberg equations of motion for $\Phi_\alpha$ and $\Pi_\alpha$ are 
\begin{eqnarray}
\label{y1}
[iH, \Phi_\alpha] & = &  \partial_t \Phi_\alpha \ = \   \Pi_\alpha \\
\label{y2}
[iH, \Pi_\alpha ] & = & \partial_t \Pi_\alpha \ = \ \partial_x\partial_x \Phi_\alpha -\mass^2\Phi_\alpha\ ,
\end{eqnarray}
which imply that $\Phi_\alpha$ satisfy the Klein--Gordon equation
\beq
\label{eq.KG}
\partial_t\partial_t \Phi_\alpha-\partial_x\partial_x \Phi_\alpha + \mass^2\Phi_\alpha=0\ .
\eeq

The Lagrangian for $\Phi_\alpha$ is 
\beq
\label{eq.LPhi}
L  =  \int \intd^3 x\, \mc{L} =  \frac{1}{2}
\int \intd^3 x\, \big[(\partial_t\Phi_\alpha) (\partial_t\Phi_\beta) \epsilon^{\alpha\beta}  - (\partial_x\Phi_\alpha) (\partial_x\Phi_\beta) \epsilon^{\alpha\beta}  - 
\mass^2\Phi_\alpha\Phi_\beta \epsilon^{\alpha\beta}\big]\ ,
\eeq
where $\Phi_\alpha$ is a real valued field in the bosonic case and a  
real Grassmann-algebra valued field in the fermionic case. 
The Euler--Lagrange equation corresponding to (\ref{eq.LPhi}) is the Klein--Gordon equation above.

The canonical momentum for $\Phi_\alpha$ is  
$
\tilde{\Pi}^\alpha= \frac{\partial \mc{L}}{\partial (\partial_t \Phi_\alpha)}=\epsilon^{\alpha\beta} \partial_t \Phi_\beta
$,
where $\mc{L}$ denotes the Lagrangian density.
The canonical equal time (anti)commutation relations are then
$
[ \Phi_\alpha(x,t),\tilde{\Pi}^\beta(x',t) ] =  \ii {\delta_\alpha}^\beta \delta^3(x-x')
$
in the bosonic case, and 
$
\{ \Phi_\alpha(x,t),\tilde{\Pi}^\beta(x',t) \}  =   -\ii {\delta_\alpha}^\beta \delta^3(x-x')
$
in the fermionic case.
The relation between $\tilde{\Pi}^\alpha$ and $\Pi_\alpha$ 
is
$
\Pi_\alpha = \epsilon_{\alpha\beta}\tilde{\Pi}^\beta
$.

Formulas for Feynman propagators, Green functions and (anti)commutation relations of the multi-component Klein--Gordon fields can be found in Appendix \ref{sec.feynman}.

The simplest way to construct a complex field $\Psi_\alpha$ 
is to combine two identical copies $\Phi_{1\alpha}$, $\Phi_{2\alpha}$,
of a real field as
$\Psi_\alpha = \frac{1}{\sqrt{2}}(\Phi_{1 \alpha}+\ii \Phi_{2 \alpha})$, assuming that 
$\Phi_{1\alpha}$ and $\Phi_{2\alpha}$ (anti)commute.
In this case the Hamiltonian operator is the sum of the Hamiltonian operators of 
$\Phi_{1\alpha}$ and $\Phi_{2\alpha}$, and the same applies to the Lagrangian.
In terms of $\Psi_\alpha$, they take the form
\begin{eqnarray}
\label{eq.Hc}
H & = &  \int \intd^3 x\, :\big[ \Pi_\alpha^\dagger \Pi_\beta \epsilon^{\alpha\beta}  + (\partial_x\Psi_\alpha^\dagger) (\partial_x\Psi_\beta) \epsilon^{\alpha\beta}  + \mass^2\Psi_\alpha^\dagger\Psi_\beta \epsilon^{\alpha\beta} \big]:\ ,
\end{eqnarray}
where  $\Pi_\alpha= \frac{1}{\sqrt{2}}(\Pi_{1\alpha}+  \ii\Pi_{2\alpha})$,
and 
\beq
\label{eq.LPsi}
L  =  \int \intd^3 x\, \mc{L} =  
\int \intd^3 x\, \big[(\partial_t\Psi_\alpha^\dagger) (\partial_t\Psi_\beta) \epsilon^{\alpha\beta}  - (\partial_x\Psi_\alpha^\dagger) (\partial_x\Psi_\beta) \epsilon^{\alpha\beta}  - 
\mass^2\Psi_\alpha^\dagger\Psi_\beta \epsilon^{\alpha\beta}\big]\ .
\eeq
In the bosonic case the dagger ${}^\dagger$ denotes simple complex conjugation here, 
but to fermionic quantities the rule $(\varphi\chi)^\dagger=\chi^\dagger \varphi^\dagger$ applies
instead of $(\varphi\chi)^\dagger=\varphi^\dagger \chi^\dagger$.

It should be noted that the self-adjointness of (\ref{eq.Hc}) or the reality of 
(\ref{eq.LPsi}) allows any Hermitian $\epsilon^{\alpha\beta}$ matrix, 
and neither the symmetric nor the antisymmetric part of $\epsilon^{\alpha\beta}$
cancels out from (\ref{eq.Hc}) and (\ref{eq.LPsi}). Thus it is possible to consider
complex fields with $\epsilon^{\alpha\beta}$ matrices that are Hermitian but 
do not necessarily satisfy the more restrictive conditions 
(\ref{eq.eps1})-(\ref{eq.eps4}). 
Also in this general case 
one can replace $\Psi$ by a $2N$-component real field that has
$\Phi_1$ as the first $N$ components and $\Phi_2$ as the second $N$ components, where 
$\Phi_1$ and $\Phi_2$ are the real and imaginary parts of $\Psi$ normalized so that 
$\Psi_\alpha=\frac{1}{\sqrt{2}}(\Phi_{1\alpha}+\ii\Phi_{2\alpha})$. In terms of this $2N$-component real field, the Hamiltonian operator and the Lagrangian take the form (\ref{eq.H}) and
(\ref{eq.LPhi}) with an $\epsilon$ that can be written in block matrix form as
\beq
\label{eq.ar1}
\left(
\begin{array}{rr}
\epsilon_1 & -\epsilon_2 \\
\epsilon_2 & \epsilon_1
\end{array}
\right)
\eeq
in the bosonic case and as
\beq
\label{eq.ar2}
\left(
\begin{array}{rr}
\ii\epsilon_2 & \ii\epsilon_1 \\
-\ii\epsilon_1 & \ii\epsilon_2
\end{array}
\right)
\eeq
in the fermionic case, where $\epsilon_1$ and $\epsilon_2$ are the real and imaginary parts
of $\epsilon$ so that $\epsilon^{\alpha\beta}=\epsilon_1^{\alpha\beta}+\ii\epsilon_2^{\alpha\beta}$. 
(\ref{eq.ar1}) and (\ref{eq.ar2}) also satisfy (\ref{eq.eps1})-(\ref{eq.eps4}). 
These considerations show that the real and imaginary parts of $\Psi$ (anti)commute only 
if $\epsilon_2=0$ in the bosonic case and $\epsilon_1=0$ in the fermionic case, i.e.\ if $\epsilon^{\alpha\beta}$ satisfy (\ref{eq.eps1})-(\ref{eq.eps4}).

Complex fields with $\epsilon$ matrices that are symmetric and real if the field is 
fermionic and antisymmetric and imaginary if the field is bosonic will have relevance 
in Section \ref{sec.RSS}, where the fields with reversed spin-statistics relation are described.
Such complex fields will be denoted by $\Upsilon$ in Section \ref{sec.RSS}.

\subsubsection{Fock space}
\label{sec.fock}

The states in the Fock space are created from the vacuum $\ket{0}$ by the operators $a_\alpha^\dagger(k)$; in particular
states with polarization vector $p^\alpha$ and momentum $k_\mu$ are created by $p^\alpha a_\alpha^\dagger(k)$ and annihilated by 
$p^{\alpha *} a_\alpha(k)$.
Polarization vectors have $N$ components that can take arbitrary complex values.

On the space of polarization vectors, which we denote by $\hat{V}$, we introduce 
the scalar product 
\beq
\braketv{p_1}{p_2}=(p_1)^{\alpha *}(p_2)^\beta \epsilon_{\alpha\beta}\ .
\label{eq.sp00}
\eeq 
Here the ${}^*$ denotes componentwise complex conjugation.
On the dual space, denoted by $V$, we introduce the scalar product
\beq
\braketv{p_1}{p_2}=(p_1)_\alpha^*(p_2)_\beta \epsilon^{\beta\alpha}\ .
\label{eq.sp01}
\eeq
With these definitions, if $\hat{u}_i$, $i=1,\dots, N$, is a basis in $\hat{V}$ such that
$\braketv{\hat{u}_i}{\hat{u}_j}=s_i\delta_{ij}$, where $s_i$ is either $1$ or $-1$, 
then the basis $u_i$ dual to $\hat{u}_i$ also satisfies 
$\braketv{u_i}{u_j}=s_i\delta_{ij}$, and vice versa.
(The dual basis is defined by the property $(\hat{u}_i)^\alpha (u_j)_\alpha = \delta_{ij}$.)  
The elements of $V$ are called dual polarization vectors.

The (anti)commutation relation of the creation and annihilation operators for particles in general polarization states
is 
\beq
\label{eq.gp1}
[(p_1)^{\alpha *} a_\alpha(k_1) , (p_2)^\beta a_\beta^\dagger(k_2) ]_\pm  =  \braketv{p_1}{p_2} \delta^3(k_1-k_2)\omega(k_1)\ .
\eeq
This shows that the creation and annihilation operators of particles that have orthogonal polarizations 
with respect to $\braketv{\ }{\ }$
(anti)commute.

The scalar product on one-particle states is
\beq
\label{eq.prod}
\braket{k,p}{k',p'}=
\braketv{p}{p'}
\delta^3(k-k')
\omega(k)\ .
\eeq
It is straightforward to extend this formula to multi-particle states; see Ref. \cite{toth}.

The eigenvalue of $H$ on 
a multi-particle state containing particles with momenta $k_1$, $k_2$, \dots, $k_j$ is 
$\omega(k_1)+\omega(k_2)+ \dots +\omega(k_j)$,
in particular the spectrum of $H$ is non-negative.

As can be seen from (\ref{eq.prod}), 
the definiteness properties of the scalar product on the Hilbert space defined above are determined by the 
signature of the scalar product $\braketv{\ }{\ }$ given in (\ref{eq.sp00}).
In the case of fermions, $\braketv{\ }{\ }$ has the signature
$(N/2,N/2)$, whereas 
in the case of bosons $\braketv{\ }{\ }$ has the same signature
as $\epsilon_{\alpha\beta}$.

\subsection{Lorentz transformation properties}
\label{sec.lorentz}

For simplicity, in the previous sections the auxiliary multi-component Klein--Gordon fields
were introduced without defining Lorentz transformation properties 
for them. We proceed now by defining Lorentz-covariant auxiliary multi-component Klein--Gordon fields.  

The definition of a Lorentz-covariant multi-component Klein--Gordon field $\Phi$ is as follows.\\[1mm]
1) Let $D$ be a finite dimensional real representation of the group $SL(2,\CC)$.
In the bosonic (commuting) case, $D$ should be a bosonic (integer spin) representation, whereas in the fermionic (anticommuting)
case $D$ should be fermionic (half-integer spin). A representation $D$ is bosonic or fermionic if its decomposition into 
irreducible representations contains only bosonic or fermionic, respectively, components.
An irreducible representation $(n/2, m/2)$, $n,m\in\ZZ$, $n,m\ge 0$, is bosonic if $n+m$ is even and fermionic if
$n+m$ is odd. $D$ is the representation according to which $\Phi$ will transform.\\   
2) An invariant complex conjugation in $D$ should be chosen; the reality of $D$ ensures, by definition, that this is possible.\\
3) An arbitrary basis consisting of real vectors in $D$ should also be chosen; the components $\Phi_\alpha$ 
are considered as vector components of $\Phi$ with respect to this basis. Thus $\Phi$ has $\dim(D)$ components. 
In this basis the invariant complex conjugation coincides with componentwise complex conjugation. \\
4) Further, an invariant tensor $\epsilon^{\alpha\beta}$
should be chosen so that it have the properties 
required of the $\epsilon^{\alpha\beta}$ tensor used in the previous sections (these properties were described at the beginning of 
Section \ref{sec.mkg}). The properties of $D$ ensure that such an  $\epsilon^{\alpha\beta}$ tensor exists (see some more details below). 
This tensor will serve as the $\epsilon^{\alpha\beta}$ tensor that appears in the general definition of
multi-component Klein--Gordon fields.\\
5) The $SL(2,\CC)$ transformation rule for $\Phi$ takes the usual form  
\beq
\label{eq.lor}
U[\Lambda]^{-1}\Phi_\alpha(x) U[\Lambda]  = 
{(\Lambda_D)_\alpha}^\beta \Phi_\beta (\Lambda_M^{-1}x)\ ,
\eeq
where $\Lambda$ is an element of $SL(2,\CC)$,
$U[\Lambda]$ is the unitary operator that represents $\Lambda$ in the Hilbert space, 
$\Lambda_D$ is the matrix that represents $\Lambda$ 
in $D$, and $\Lambda_M$ is the matrix that represents $\Lambda$ in Minkowski space-time. \\
6) The vacuum $\ket{0}$ is invariant under $SL(2,\CC)$ transformations, i.e.\ $U[\Lambda]\ket{0}=\ket{0}$.

\vspace{1mm}
It follows from (\ref{eq.lor}) that $SL(2,\CC)$ transformations act on the creation operators $a_\alpha^\dagger(k)$ as
\beq
U[\Lambda]^{-1} a_\alpha^\dagger(k) U[\Lambda]={(\Lambda_D)_\alpha}^\beta a_\beta^\dagger(\Lambda_M^T k)\ ,
\eeq 
and $a_\alpha(k)$ also has the same transformation property.
In this formula the $T$ in the superscript denotes transposition, and  
by writing $\Lambda_M^T k$ it is meant that  
$\Lambda_M^T$ acts on the dual four-vector $k_\mu=(\omega(k),k)$.

As in Ref. \cite{toth}, we shall concentrate mainly on the cases 
$D=D^{(n,m)}$ and $D=\tilde{D}^{(n)}$, where $n,m\in\ZZ$ and 
\begin{eqnarray}
D^{(n,m)} & = & (n/2,m/2)\oplus (m/2,n/2),  \qquad n > m\ge 0 \ ,  \\
\tilde{D}^{(n)} & = & (n/2,n/2),  \hspace{3.15cm}  n\ge 0\ ,
\end{eqnarray}
nevertheless other representations could be treated similarly to these cases.

$D^{(n,m)}$ and $\tilde{D}^{(n)}$ are the irreducible real
representations of $SL(2,\CC)$, therefore they are the simplest possible choices for $D$.
Any other finite dimensional real representation can be decomposed
into a direct sum of these representations.  
$D^{(1,0)}= (1/2,0)\oplus(0,1/2)$ is the Dirac representation and $\tilde{D}^{(1)}=(1/2,1/2)$ is the usual vector representation. 

For and $n,m\ge 0$, $n,m\in \ZZ$, the complex conjugate of the representation $(n/2,m/2)$ is 
equivalent to $(m/2,n/2)$,  whereas 
the dual of $(n/2,m/2)$ is equivalent to itself. 

The invariant  
complex conjugation in $\tilde{D}^{(n)}$ is unique 
up to multiplication by a complex number that has absolute value $1$. 
Regarding $D^{(n,m)}$, any invariant complex conjugation takes the subspace $(n/2,m/2)$
into $(m/2,n/2)$ and $(m/2,n/2)$ into $(n/2,m/2)$, and thus takes the two by two matrix form
\beq
\left(\begin{array}{cc}
 0 & J^{-1} \\
 J & 0
\end{array}\right)
\eeq
with respect to the decomposition $(n/2,m/2)\oplus (m/2,n/2)$. 
The complete set of invariant complex conjugations can be parametrized by a single complex number as
\beq
\left(\begin{array}{cc}
 0 & \frac{1}{\alpha^*} J^{-1} \\
 \alpha J & 0
\end{array}\right),\qquad \alpha\in \CC\ ,\quad \alpha\ne 0\ ,
\eeq
where $J$ is one fixed but otherwise freely chosen invariant complex conjugation from $(n/2,m/2)$ to $(m/2,n/2)$.

$(n/2,m/2)$ (allowing $n=m$) admits a unique (up to multiplication by a complex number) invariant bilinear form, which is nondegenerate and symmetric if $n+m$ is even and 
nondegenerate and antisymmetric if $n+m$ is odd.
If $n\ne m$, then the decomposition of $(n/2,m/2)\otimes (m/2,n/2)$ into irreducible representations
does not contain the trivial representation $(0,0)$, thus any invariant bilinear form $\epsilon$ on 
$D^{(n,m)}$ can be written in a two by two matrix form
\beq
\label{eq.epsmatr}
\left(\begin{array}{cc}
 \epsilon^L & 0 \\
 0 & \epsilon^R
\end{array}\right)
\eeq
with respect to the decomposition $(n/2,m/2)\oplus (m/2,n/2)$, 
where $\epsilon^L$ is the restriction of $\epsilon$ to $(n/2,m/2)$ and $\epsilon^R$ is the restriction of $\epsilon$ to $(m/2,n/2)$. It is also obvious that with an arbitrary choice of 
invariant bilinear forms $\epsilon^L$ and $\epsilon^R$ on $(n/2,m/2)$ and $(m/2,n/2)$, (\ref{eq.epsmatr})
is an invariant bilinear form on $D^{(n,m)}$.
Assuming that an invariant complex conjugation has been fixed, $\epsilon$ is real if
$\epsilon^R=\epsilon^{L*}$ and imaginary if $\epsilon^R=-\epsilon^{L*}$.
(Without using basis vectors, the complex conjugate of a bilinear form $\epsilon$ can be defined
via the equation $\epsilon^*(u^*,v^*)=\epsilon(u,v)^*$, where $u$ and $v$ are arbitrary vectors.)
Thus the complete set of invariant real bilinear forms on $D^{(n,m)}$ is
\beq
\left(\begin{array}{cc}
 \alpha\epsilon^L & 0 \\
 0 & \alpha^*\epsilon^{L*}
\end{array}\right),\qquad \alpha\in \CC,
\eeq
where $\epsilon^L$ is a
non-zero but otherwise arbitrarily chosen fixed bilinear form on $(n/2,m/2)$, 
and similarly the set of imaginary bilinear forms is
\beq
\left(\begin{array}{cc}
 \alpha\epsilon^L & 0 \\
 0 & -\alpha^*\epsilon^{L*}
\end{array}\right),\qquad \alpha\in \CC.
\eeq
The real invariant bilinear forms on $\tilde{D}^{(n)}$ are unique up to multiplication by a real number. 
It is also clear that any nondegenerate bilinear form on $D^{(n,m)}$ or on $\tilde{D}^{(n)}$
can be made real or imaginary by a suitable choice of the invariant complex conjugation.

For the constructions described in the present paper all choices of the invariant complex conjugation
and of $\epsilon^{\alpha\beta}$ are equivalent in the cases $D=D^{(n,m)}$, 
whereas if $D=\tilde{D}^{(n)}$, then
two slightly different cases can be distinguished (see Section \ref{sec.polar}) depending on the sign of $\epsilon^{\alpha\beta}$.

The number of the 
components of $\Phi$ is $\dim(D^{(n,m)})= 2(n+1)(m+1)$ 
if $D=D^{(n,m)}$ 
and $\dim(\tilde{D}^{(n)}) = (n+1)^2$ if $D=\tilde{D}^{(n)}$.
In the case of $D=D^{(n,m)}$, $\Phi$ is bosonic if $n+m$ is even 
and fermionic if $n+m$ is odd. In the case of $D=\tilde{D}^{(n)}$,  $\Phi$ is bosonic for any value of $n$.

\subsection{Particle spectrum}
\label{sec.polar}

In order to determine what kind of particles are described by $\Phi$, the space of allowed polarization vectors
for any momentum $k$, which is $D^*$, should be decomposed into irreducible representations of the $SU(2)$ little group that 
leaves the momentum four-vector $(\omega(k),k)$ invariant. Along with $D^*$ it is worth considering $D$ as well.
Due to Lorentz symmetry it is sufficient to focus first on $k=0$, and then obtain the decomposition corresponding to
$k\ne 0$ by boost.

In the following we
continue with the case of $D=D^{(n,m)}$, 
and discuss the case of $D=\tilde{D}^{(n)}$ subsequently.\\

\noindent
{\bf $\mathbf{D=D^{(n,m)}}$:}

For any choice of the invariant complex conjugation and of the invariant tensor
$\epsilon^{\alpha\beta}$, 
in the representations $D^{(n,m)}$ it is possible to find basis vectors  
$u_{(s)i}$, $v_{(s)i}$; $s=(n-m)/2,(n-m)/2 + 1,\dots, (n+m)/2$; $i=1,\dots,d_s$; $d_s=2s+1$; so that 
the orthogonality relations 
\beq
\label{eq.ortho1}
\braketv{u_{(s)i}}{u_{(s')j}}=\delta_{ss'}\delta_{ij} \qquad 
\braketv{v_{(s)i}}{v_{(s')j}}=-\delta_{ss'}\delta_{ij} \qquad
\braketv{u_{(s)i}}{v_{(s')j}}=0
\eeq
hold, 
where $\braketv{\ }{\ }$ denotes the scalar product introduced in Section  \ref{sec.fock}.
These basis vectors also have the property that for any fixed $s$, 
$u_{(s)i}$, $i=1,\dots,d_s$ and $v_{(s)i}$, $i=1,\dots,d_s$ span two subspaces of
$D^{(n,m)}$ that are irreducible spin $s$ representations with respect to the rotation ($SU(2)$) little group
that leaves the momentum covector $(\mass,0)$ invariant. (See Appendix \ref{sec.appb} for details on the derivation of these statements.)

The vectors $u_{(s)i}(k)$, $v_{(s)i}(k)$ are defined now as
\begin{eqnarray}
\label{eq.uik}
u_{(s)i}(k) & = &  \Lambda_D(k)u_{(s)i}\ , \qquad i=1,\dots, d_s\\
\label{eq.vik}
v_{(s)i}(k) & = & \Lambda_D(k)v_{(s)i}\ , \qquad i=1,\dots, d_s\ ,
\end{eqnarray}
where $\Lambda_D(k)$ represents in $D$ 
the unique $SL(2,\CC)$ element $\Lambda(k)$ determined by the properties that 
$\Lambda(k)$ is a continuous function of $k$, $\Lambda(0)=I$, and the Lorentz transformation
corresponding to $\Lambda(k)$ is the Lorentz boost that
takes the dual four-vector $(\mass,0)$ to $(\omega(k),k)$. 

For any fixed $s$, 
$u_{(s)i}(k)$, $i=1,\dots,d_s$, and $v_{(s)i}(k)$, $i=1,\dots,d_s$, span two subspaces of
$D^{(n,m)}$ that are (irreducible) spin $s$ representations with respect to the rotation ($SU(2)$) little group
that leaves the momentum covector $(\omega(k),k)$ invariant. 

The vectors dual to $u_{(s)i}(k)$, $v_{(s)i}(k)$ are denoted by $\hat{u}_{(s)i}(k)$, $\hat{v}_{(s)i}(k)$.
These vectors also satisfy the orthogonality relations
\beq
\label{eq.ortho2}
\braketv{\hat{u}_{(s)i}(k)}{\hat{u}_{(s')j}(k)}=\delta_{ss'}\delta_{ij} \qquad 
\braketv{\hat{v}_{(s)i}(k)}{\hat{v}_{(s')j}(k)}=-\delta_{ss'}\delta_{ij} 
\eeq
\beq
\label{eq.ortho2b}
\braketv{\hat{u}_{(s)i}(k)}{\hat{v}_{(s')j}(k)}=0\ ,
\eeq
where $\braketv{\ }{\ }$ denotes again the scalar product introduced in Section  \ref{sec.fock}, thus they 
form a complete set of orthogonal polarization vectors.
For any fixed $s$, 
$\hat{u}_{(s)i}(k)$, $i=1,\dots,d_s$, and $\hat{v}_{(s)i}(k)$, $i=1,\dots,d_s$, span two subspaces of
$D^{(n,m)*}$ that are irreducible spin $s$ representations with respect to the $SU(2)$ little group
that leaves the momentum covector $(\omega(k),k)$ invariant. 
This follows from the duality between  $u_{(s)i}(k)$, $v_{(s)i}(k)$ and $\hat{u}_{(s)i}(k)$, $\hat{v}_{(s)i}(k)$, and
is in accordance with the fact that for any $k$ the decomposition of both $D^{(n,m)*}$ and $D^{(n,m)}$ into irreducible representations of
the corresponding $SU(2)$ little group is 
\beq
\bigg(\frac{n+m}{2}\bigg)\oplus \bigg(\frac{n+m}{2}\bigg) \oplus \bigg(\frac{n+m}{2}-1\bigg)\oplus \bigg(\frac{n+m}{2}-1\bigg) \oplus \dots \oplus 
\bigg(\frac{n-m}{2}\bigg)\oplus \bigg(\frac{n-m}{2}\bigg)\ .
\eeq

The following completeness relations can be written down for 
$u_{(s)i}(k)$, $v_{(s)i}(k)$, $\hat{u}_{(s)i}(k)$, $\hat{v}_{(s)i}(k)$:
\begin{eqnarray}
\label{eq.c1}
{\delta_\alpha}^\beta & = &
\sum_s \sum_{i=1}^{d_s} u_{(s)i\alpha}(k)\hat{u}_{(s)i}^\beta(k) + \sum_s\sum_{i=1}^{d_s} v_{(s)i\alpha}(k)\hat{v}_{(s)i}^\beta(k)\\
\label{eq.c2}
{\delta_\alpha}^\beta & = &
\sum_s\sum_{i=1}^{d_s} u_{(s)i\alpha}(k)^*\hat{u}_{(s)i}^\beta(k)^* + \sum_s\sum_{i=1}^{d_s} v_{(s)i\alpha}(k)^*\hat{v}_{(s)i}^\beta(k)^* \ .
\end{eqnarray}
(\ref{eq.c2}) is obtained from (\ref{eq.c1}) by complex conjugation. 

The orthogonality relations (\ref{eq.ortho1}) imply that 
\beq
\hat{u}_{(s)i}^\beta(k) = \epsilon^{\beta\alpha} u_{(s)i\alpha}(k)^*\ ,\qquad 
\hat{v}_{(s)i}^\beta(k) = -\epsilon^{\beta\alpha} v_{(s)i\alpha}(k)^*\ ,
\eeq 
thus (\ref{eq.c1}) can also be written in the form
\beq
\label{eq.c1x}
{\delta_\alpha}^\beta \ = \
\sum_s \sum_{i=1}^{d_s} u_{(s)i\alpha}(k)\bar{u}_{(s)i}^\beta(k) - \sum_s\sum_{i=1}^{d_s} v_{(s)i\alpha}(k)\bar{v}_{(s)i}^\beta(k)\ ,\\
\eeq
where the notation
\beq
\bar{u}^\alpha = \epsilon^{\alpha\beta}u_\beta^*
\eeq
is used. In the case of the Dirac field, $\bar{u}$ is $-1$ times the Dirac-conjugate of $u$ (see Section \ref{sec.dirac}), 
therefore (\ref{eq.c1x}) has a form that is more familiar from textbooks than (\ref{eq.c1}). 
However, (\ref{eq.c1}) is more general than (\ref{eq.c1x}), since it does not make use of the orthogonality properties 
of $u_{(s)i\alpha}(k)$ and $v_{(s)i\alpha}(k)$.

The creation operators that create particles with polarizations $\hat{u}_{(s)i}(k)$ and $\hat{v}_{(s)i}(k)$
are
\begin{eqnarray}
c_{(s)i}^\dagger(k) & = & \hat{u}_{(s)i}(k)^\alpha a_\alpha^\dagger(k),\qquad i=1,\dots, d_s\\
f_{(s)i}^\dagger(k) & = & \hat{v}_{(s)i}(k)^\alpha a_\alpha^\dagger(k),\qquad i=1,\dots, d_s\ .
\end{eqnarray}
$c_{(s)i}^\dagger(k)$ is related to $c_{(s)i}^\dagger(0)$ by the formula
\beq
\label{eq.57}
c_{(s)i}^\dagger(k)  = U[\Lambda(k)]    c_{(s)i}^\dagger(0)  U[\Lambda(k)]^{-1}\ ,
\eeq
where $\Lambda(k)$ is the boost defined as above.    
The same formula applies also to $f_{(s)i}^\dagger(k)$.

It follows from (\ref{eq.gp1}) and (\ref{eq.ortho2}), (\ref{eq.ortho2b}) that
the operators $c_i(k)$ and $f_i(k)$ satisfy the (anti)commutation relations
\begin{eqnarray}
&& [ c_{(s)i}(k),c_{(s')j}^\dagger (k') ]_\pm = \delta_{ss'}\delta_{ij}\delta^3(k-k')\omega(k)\\
&& [ f_{(s)i}(k),f_{(s')j}^\dagger (k') ]_\pm = -\delta_{ss'}\delta_{ij}\delta^3(k-k')\omega(k),
\end{eqnarray}
all other (anti)commutators of them are zero.
This shows that the one-particle states created by 
$c_{(s)i}^\dagger(k)$ have positive scalar product with themselves,
whereas 
the one-particle states created by 
$f_{(s)i}^\dagger(k)$ have negative scalar product with themselves.

By making use of relations (\ref{eq.c1}) and (\ref{eq.c2}), $\Phi$ can be expressed in terms of 
$c_{(s)i}$ and $f_{(s)i}$ as
\begin{eqnarray}
&& \Phi_\alpha (x,t)= \nonumber \\
&& \hspace{1cm} \int \frac{\intd^3 k}{\sqrt{2}(\sqrt{2\pi})^3 \omega(k)} \big[e^{\ii kx}e^{\ii \omega(k)t}
    \sum_s  \sum_{i=1}^{d_s}  \{ u_{(s)i\alpha}(k) c_{(s)i}^\dagger(k) + v_{(s)i\alpha}(k) f_{(s)i}^\dagger(k) \} \nonumber \\
&& \hspace{1.1cm} +e^{-\ii kx}e^{-\ii\omega(k)t}
    \sum_s \sum_{i=1}^{d_s} \{ u_{(s)i\alpha}(k)^* c_{(s)i}(k) +v_{(s)i\alpha}(k)^* f_{(s)i}(k) \} \big]\ .
\label{eq.64}
\end{eqnarray}
The values of $s$ over which the summation is done here are $(n-m)/2,(n-m)/2+1,\dots,(n+m)/2$.
This formula, along with the properties of the polarization vectors discussed above, 
shows that $\Phi$ describes $2$ kinds of particles of mass $\mu$ for any value of $s$. 
One of them is physical and one is non-physical (in the sense that the one-particle states have negative 
scalar product with themselves).

The Hamiltonian operator can be expressed as
\beq
H=
\int \frac{\intd^3 k}{\omega(k)}\ \omega(k) \sum_s \sum_{i=1}^{d_s} \big[c_{(s)i}^\dagger(k) c_{(s)i}(k) 
- f_{(s)i}^\dagger(k) f_{(s)i}(k) \big] \ .
\eeq

\noindent
{\bf $\mathbf{D=\tilde{D}^{(n)}}$:}

The case of  $D=\tilde{D}^{(n)}$  is very similar to the case of $D=D^{(n,m)}$; the main difference is that 
the basis vectors 
$u_{(s)i}$ are defined now only for $s\in S_+=\{ n,n-2,n-4,\dots \}$, whereas 
$v_{(s)i}$ are defined only for $s\in S_-=\{ n-1,n-3,n-5,\dots \}$. 
(Here a particular choice of the sign of $\epsilon^{\alpha\beta}$ is assumed;
$S_+$ and $S_-$ will be interchanged if the sign of $\epsilon^{\alpha\beta}$ is changed.)
The decomposition of $\tilde{D}^{(n)}$ and $\tilde{D}^{(n)*}$ into 
irreducible representations of 
the $SU(2)$ little group corresponding to any momentum $k$ is 
\beq
(n)\oplus (n-1) \oplus (n-2)\oplus \dots \oplus 
(1)\oplus (0)\ .
\eeq

The completeness relations have to be modified in the following obvious way:
\begin{eqnarray}
\label{eq.c3}
{\delta_\alpha}^\beta & = &
\sum_{s\in S_+} \sum_{i=1}^{d_s} u_{(s)i\alpha}(k)\hat{u}_{(s)i}^\beta(k) + \sum_{s\in S_-} \sum_{i=1}^{d_s} v_{(s)i\alpha}(k)\hat{v}_{(s)i}^\beta(k)\\
\label{eq.c4}
{\delta_\alpha}^\beta & = &
\sum_{s\in S_+} \sum_{i=1}^{d_s} u_{(s)i\alpha}(k)^*\hat{u}_{(s)i}^\beta(k)^* + \sum_{s\in S_-} \sum_{i=1}^{d_s} v_{(s)i\alpha}(k)^*\hat{v}_{(s)i}^\beta(k)^* \ .
\end{eqnarray}

The mode expansion of $\Phi$ takes the form
\beq
\Phi=\phi + \phi^-\ ,
\eeq 
where  
\begin{eqnarray}
&& \phi_\alpha (x,t)=\int \frac{\intd^3 k}{\sqrt{2}(\sqrt{2\pi})^3 \omega(k)} \big[ e^{\ii kx}e^{\ii \omega(k)t}
    \sum_{s\in S_+}  \sum_{i=1}^{d_s}   u_{(s)i\alpha}(k) c_{(s)i}^\dagger(k)   \nonumber \\
&& \hspace{3cm} +e^{-\ii kx}e^{-\ii\omega(k)t}
    \sum_{s\in S_+} \sum_{i=1}^{d_s}  u_{(s)i\alpha}(k)^* c_{(s)i}(k)   \big]
\end{eqnarray}
and 
\begin{eqnarray}
&& (\phi^-)_\alpha (x,t)=\int \frac{\intd^3 k}{\sqrt{2}(\sqrt{2\pi})^3 \omega(k)} \big[e^{\ii kx}e^{\ii \omega(k)t}
    \sum_{s\in S_-}  \sum_{i=1}^{d_s}  v_{(s)i\alpha}(k) f_{(s)i}^\dagger(k)  \nonumber \\
&& \hspace{3cm} +e^{-\ii kx}e^{-\ii\omega(k)t}
    \sum_{s\in S_-} \sum_{i=1}^{d_s}  v_{(s)i\alpha}(k)^* f_{(s)i}(k)  \big]\ .
\end{eqnarray}
In the present case 
$\Phi$ describes one kind of physical particles of mass $\mu$ for any value of $s\in S_+$ and 
one kind of non-physical particles of mass $\mu$ for any value of $s\in S_-$ 
(again, the non-physical nature is understood to mean that the one-particle states have negative 
scalar product with themselves). 
In particular, $\phi$ and $\phi^\dagger$ create the physical particles and $\phi^-$ and $\phi^{-\dagger}$ the non-physical ones.

The Hamiltonian operator can be written as
\begin{eqnarray}
 H & = &
\int \frac{\intd^3 k}{\omega(k)}\ \omega(k) \sum_{s\in S_+} \sum_{i=1}^{d_s} \big[c_{(s)i}^\dagger(k) c_{(s)i}(k)  \big] \nonumber \\
&& \hspace{1cm} - \int \frac{\intd^3 k}{\omega(k)}\ \omega(k) 
\sum_{s\in S_-} \sum_{i=1}^{d_s}  \big[f_{(s)i}^\dagger(k) f_{(s)i}(k)\big] \ .
\end{eqnarray}

In the following, polarization vectors $\hat{u}_{(s)i}(k)$, $\hat{v}_{(s)i}(k)$ having the properties described in this section  
will be called standard polarization vectors.
It should also be noted that these standard polarization vectors are usually chosen so that they have the additional property
of being eigenstates 
of the $SU(2)$ little group generator $\Lambda_{D^*}(k)M_{3D^*}\Lambda_{D^*}(k)^{-1}$.

\subsection{Elementary higher spin fields}
\label{sec.hsf}

In this section the last step in the construction of higher spin fields is described.
We continue to focus on the representations $D=D^{(n,m)}$ and $D=\tilde{D}^{(n)}$, treating these cases together.

As was seen in Section \ref{sec.polar}, the fields $\Phi$ can create states that have negative 
scalar product with themselves. Moreover, $\Phi$ can generally create states with  
several different spins, except in the cases $D=D^{(n,0)}$.
An elementary real higher spin field $\phi_{(s)}$ that creates states with a specific spin $s$ and
having positive scalar product with themselves 
can be obtained from $\Phi$ by applying to it a 
suitable real Poincar\'e-invariant differential operator
$\mc{D}_{(s)}[\partial]$, which projects out the desired modes: 
\beq
\label{eq.projection}
\phi_{(s)}=\mc{D}_{(s)}[\partial]\Phi\ .
\eeq
These $\mc{D}_{(s)}[\partial]$ operators are orthogonal 
projections in the space of the solutions of the Klein--Gordon equation
(\ref{eq.KG}), i.e.\ $\mc{D}_{(s)}[\partial]\mc{D}_{(s)}[\partial]\Phi =\mc{D}_{(s)}[\partial]\Phi$ and
$\mc{D}_{(s)}[\partial]\mc{D}_{(s')}[\partial]\Phi =0$, $s\ne s'$, hold.
Thus $\phi_{(s)}$ satisfies in addition to the Klein--Gordon equation 
\beq
(\partial_t^2-\partial_x^2+\mass^2)\phi_{(s)}=0
\eeq
the differential equations
\begin{eqnarray}
\label{eq.de1}
\mc{D}_{(s)}[\partial] \phi_{(s)} & = &  \phi_{(s)} \\
\label{eq.de2}
\mc{D}_{(s')}[\partial] \phi_{(s)} & = & 0\ ,\qquad s'\ne s \ .
\end{eqnarray}
$\mc{D}_{(s)}[\partial]$ has the following action on the standard polarization modes of $\Phi$:
\begin{eqnarray}
\label{eq.d1}
{\mc{D}_{(s)}[\ii k]_\alpha}^\beta  u_{(s')i\beta}(k) & = & \delta_{ss'} u_{(s)i\alpha}(k) \\
\label{eq.d2}
{\mc{D}_{(s)}[\ii k]_\alpha}^\beta v_{(s')i\beta}(k) & = & 0 \\
\label{eq.d3}
{\mc{D}_{(s)}[-\ii k]_\alpha}^\beta  u_{(s')i\beta}(k)^* & = & \delta_{ss'} u_{(s)i\alpha}(k)^* \\
\label{eq.d4}
{\mc{D}_{(s)}[-\ii k]_\alpha}^\beta  v_{(s')i\beta}(k)^* & = & 0 \ .
\end{eqnarray}
Equations (\ref{eq.d3}) and (\ref{eq.d4}) are obtained by complex conjugation from (\ref{eq.d1}) and (\ref{eq.d2}), taking into consideration the reality of 
$\mc{D}_{(s)}[\partial]$.
The notation $\mc{D}_{(s)}[\ii k]$ means that $\partial_\mu$ in the expression of $\mc{D}_{(s)}[\partial]$
is replaced by the momentum covector $(\ii \omega(k),\ii k)$. $\mc{D}_{(s)}[\partial]$ is a polynomial 
of $\partial_\mu$, with coefficients that are real $SL(2,\CC)$-invariant tensors. 
The property that $\mc{D}_{(s)}[\partial]$ is an invariant differential operator implies that the Poincar\'e transformation properties
of $\phi_{(s)}$ are the same as those of $\Phi$.

A real invariant linear differential operator $\mc{D}^-[\partial]$ 
which projects out the non-physical modes
from $\Phi$ (i.e.\ that part of $\Phi$ which creates one-particle states that have negative scalar product with themselves)
can also be introduced.

$\mc{D}_{(s)}[\partial]$ and 
$\mc{D}^-[\partial]$ form a complete set of projectors on the space of the solutions of the Klein--Gordon equation:
\beq
\sum_s {\mc{D}_{(s)}[\partial]_\alpha}^\beta \Phi_\beta + {\mc{D}^-[\partial]_\alpha}^\beta \Phi_\beta = \Phi_\alpha\ .
\eeq
$\phi_{(s)}$ also satisfies the differential
equation
\beq
\label{eq.dmp}
\mc{D}^-[\partial] \phi_{(s)} =  0 \ .
\eeq

The properties (\ref{eq.d1})-(\ref{eq.d4}) of $\mc{D}_{(s)}[\partial]$ imply that
$\phi_{(s)}$ has the mode expansion
\begin{eqnarray}
&& \phi_{(s)\alpha} (x,t)=\int \frac{\intd^3 k}{\sqrt{2}(\sqrt{2\pi})^3 \omega(k)}  \sum_{i=1}^{d_s}
\big[ e^{\ii kx}e^{\ii \omega(k)t}
        u_{(s)i\alpha}(k) c_{(s)i}^\dagger(k) \nonumber \\
&& \hspace{5cm} +e^{-\ii kx}e^{-\ii\omega(k)t}
      u_{(s)i\alpha}(k)^* c_{(s)i}(k)  \big] \ .
\label{eq.psimodex}
\end{eqnarray}
This shows that $\phi_{(s)}$ describes one particle of mass $\mass$ and spin $s$.

Generally it is an algebraic problem to find the $\mc{D}_{(s)}[\partial]$ and $\mc{D}^-[\partial]$ operators. 
Without going into details, we note that for small values of $m$ or $n$   
it is not difficult to find these operators by direct calculations, and relevant general results for any spin
can be found
in Refs. \cite{Weinberg,Weinberg3,WeinbergQFT} (see also Ref. \cite{toth}).
Explicit expressions for relevant projection operators 
for the Rarita--Schwinger fields are also available in the literature, e.g.\ in Refs. \cite{Fronsdal,BF,Peng2}.

Complex fields $\psi_{(s)}$ can be constructed in the same way as in Section \ref{sec.mkg}, i.e.\ 
by taking two identical (anti)commuting copies $\phi_{1(s)}$, $\phi_{2(s)}$ of a real field and combining them as 
$\psi_{(s)}=\frac{1}{\sqrt{2}}(\phi_{1(s)}+\ii\phi_{2(s)})$. The relation (\ref{eq.projection}) remains valid also between
the complex auxiliary field and $\psi_{(s)}$. Obviously, the set of particles described by $\psi_{(s)}$, or by the complex auxiliary field, is doubled in comparison with the real fields.

Fields transforming according to the representations $(n/2,m/2)$, $n\ne m$ (i.e.\ chiral fields) can also be obtained from
the fields that transform according to $D^{(n,m)}$ by projection on the $(n/2,m/2)$ or $(m/2,n/2)$ component.
More details on chiral fields can be found in Ref. \cite{toth}.

Formulas for Feynman propagators, Green functions and (anti)commutation relations of the elementary higher spin fields can be found in Appendix \ref{sec.feynman}.

Regarding the construction of interaction Hamiltonians from higher spin fields, we refer the reader to Refs.
\cite{WeinbergQFT,Weinberg,Peng2}, and to Ref.
\cite{toth}, especially Section 5.4.

\section{Fields of arbitrary spin with reversed spin-statistics relation}
\label{sec.RSS}

We turn now to the description of the construction of 
fields with reversed spin-statistics relation (abbreviated hereafter as RSS fields) in the framework described in Section \ref{sec.1}. 
The RSS fields will be constructed from normal fields;
more precisely,
data that specify RSS fields will be constructed from data used to specify the normal fields. 
We do not introduce extra notation to distinguish RSS fields from normal fields.  

The main ingredients in the definition of a normal higher spin field are the representation $D$, the bilinear form 
$\epsilon^{\alpha\beta}$, the invariant complex conjugation, the real basis vectors in $D$, 
the standard polarization vectors, and the projection operator $\mc{D}_{(s)}[\partial]$. 
Assuming that these objects are given, a corresponding new $SL(2,\CC)$ representation, bilinear form, complex conjugation, 
real basis vectors, standard polarization vectors and 
projection operator, which will characterize the RSS field that corresponds to the original normal field, can be defined,
as described below. The most important definitions are those of the $SL(2,\CC)$ representation (\ref{eq.drss}), 
the bilinear form (\ref{eq.etilde}), and the projection operator (\ref{eq.RSSproj}).
The RSS fields obtained in this way will fit entirely in the framework of Section \ref{sec.1}, with the only difference that bosonic RSS fields will transform 
according to half-integer spin representations of $SL(2,\CC)$ and fermionic RSS fields according to integer spin representations.

For the representation $D_{RSS}$ according to which the RSS auxiliary field $\Phi$ transforms we take
the direct sum of two copies $D_A$ and $D_B$ of the representation $D$: 
\beq
\label{eq.drss}
D_{RSS}=D_A\oplus D_B\ ,
\eeq
where the subscripts ${}_A$ and ${}_B$ are introduced to identify the two components of the direct sum.

We use the notation
\beq
U = \left( \begin{array}{c}
U_A \\
U_B \\
\end{array}\right)
\eeq
for vectors in $D_{RSS}$, i.e.\ we denote by $U_A$ and $U_B$ the projection of a vector $U$ 
on the components $D_A$ and $D_B$, respectively, of $D_{RSS}$. 
In particular, the auxiliary field $\Phi$ has the two components $\Phi_A$ and $\Phi_B$, which transform according to $D_A$ and $D_B$.
For dual vectors 
(which are in $D_{RSS}^*$, and which have, according to the definitions in Section \ref{sec.fock}, the role of polarization vectors) we use a similar notation 
\beq
\hat{U}=(\hat{U}^A,\hat{U}^B) \ .
\eeq

We define the invariant complex conjugation in $D_{RSS}$ in a straightforward way via the complex conjugation 
in the two components $D_A$ and $D_B$:
\beq
U^* = \left( \begin{array}{c}
U_A^* \\
U_B^* \\
\end{array}\right) \ .
\eeq
For dual vectors we have then
\beq
\hat{U}^*=(\hat{U}^{A*},\hat{U}^{B*})\ .
\eeq

With the above choice of the complex conjugation in $D_{RSS}$,
if $e_i$ are real basis vectors in $D$, then the vectors 
$\left( \begin{array}{c}
e_i \\
0 \\
\end{array}\right)$,
$\left( \begin{array}{c}
0 \\
e_i \\
\end{array}\right)$ are real basis vectors 
in $D_{RSS}$. This applies if,
in particular, $\{ e_i\}$ is the fixed real basis in $D$ mentioned in Section \ref{sec.lorentz},
thus the fixed real basis in $D_{RSS}$ can be chosen to be
$\left\{ \left( \begin{array}{c}
e_i \\
0 \\
\end{array}\right),
\left( \begin{array}{c}
0 \\
e_i \\
\end{array}\right)\right\}$.

For the vector components of vectors in $D_{RSS}$ we use the notation $U_\alpha$, where\\
$\alpha=1,2, \dots, 2\dim(D)$. This means
that we do not introduce a new type of index for vectors in $D_{RSS}$, nevertheless this should not cause confusion. 

For the invariant bilinear form on $D_{RSS}$ we take the tensor $\tilde{\epsilon}$ defined as 
\beq
\tilde{\epsilon}(U,V) \equiv \tilde{\epsilon}^{\alpha\beta}U_\alpha V_\beta 
= \ii[\epsilon(U_A,V_B)-\epsilon(U_B,V_A)]
\equiv \ii[\epsilon^{\alpha\beta}U_{A\alpha}V_{B\beta} - \epsilon^{\alpha\beta}U_{B\alpha}V_{A\beta}] \ ,
\eeq
where $\epsilon$ denotes the invariant bilinear form on $D$.
$\tilde{\epsilon}$ can be written in block matrix form as
\beq
\label{eq.etilde}
\tilde{\epsilon}=\left( \begin{array}{cc}
0 & \ii \epsilon \\
-\ii \epsilon & 0 \\
\end{array}\right) \ .
\eeq
If $\epsilon$ is real and symmetric, then $\tilde{\epsilon}$ is purely imaginary and antisymmetric, and vice versa. 
This implies that the normal auxiliary field, to which $D$ and $\epsilon$ belong, and the 
corresponding RSS auxiliary field have opposite commutation properties.

The inverse of  $\tilde{\epsilon}$ is given by
\begin{eqnarray}
&& \tilde{\epsilon}^{-1}(\hat{U},\hat{V}) \equiv  \tilde{\epsilon}_{\alpha\beta}\hat{U}^\alpha \hat{V}^\beta = \nonumber \\ 
&& \ii[\epsilon^{-1}(\hat{U}^A,\hat{V}^B)-\epsilon^{-1}(\hat{U}^B,\hat{V}^A)]
\equiv 
\ii[\epsilon_{\alpha\beta}\hat{U}^{A\alpha} \hat{V}^{B\beta} - \epsilon_{\alpha\beta}\hat{U}^{B\alpha}\hat{V}^{A\beta}]
\ .
\end{eqnarray}
$\tilde{\epsilon}^{-1}$ can be written in block matrix form as
\beq
\tilde{\epsilon}^{-1}=\left( \begin{array}{cc}
0 & \ii \epsilon^{-1} \\
-\ii \epsilon^{-1} & 0 \\
\end{array}\right) \ .
\eeq

The equal time (anti)commutation relations (\ref{eq.ac1}) of the RSS auxiliary field $\Phi$ and of the corresponding momentum field 
$\Pi$ can be written in terms of the components
$\Phi_A$, $\Phi_B$, $\Pi_A$, $\Pi_B$ as 
\begin{eqnarray}
\label{eq.RSScomm1}
[ \Phi_{A\alpha}(x,t) , \Pi_{B\beta} (x',t) ]_\pm & = & - \epsilon_{\alpha\beta} \delta^3(x-x') \\
{} [ \Phi_{B\alpha} (x,t) , \Pi_{A\beta} (x',t) ]_\pm & = & \phantom{-}\epsilon_{\alpha\beta} \delta^3(x-x') \ .
\end{eqnarray} 
Other (anti)commutators of $\Phi_A$, $\Phi_B$, $\Pi_A$, $\Pi_B$ are zero.

We note that $\Phi_A$ and $\Phi_B$ can be combined into a complex field
$\Upsilon$ as $\Upsilon=\frac{1}{\sqrt{2}}(\Phi_A+\ii\Phi_B)$, which has the 
Lagrangian (\ref{eq.LPsi}) with the original $\epsilon$ matrix, and which transforms
according to the representation $D$. 
This $\Upsilon$ is thus a complex field that is characterized by a real and symmetric $\epsilon$ matrix if it is fermionic and by an imaginary and antisymmetric $\epsilon$ matrix if it is 
bosonic, which is the reverse of prescription made for $\epsilon$ in Section \ref{sec.mkg}.
In other words, a real RSS auxiliary field can also be represented as a complex field
that is obtained from a normal complex auxiliary field by simply interchanging 
bosonic and fermionic commutation properties.

\subsection{Polarization vectors and particle spectrum}
\label{sec.RSSpolar}

The spectrum of particle states that can be created by $\Phi$ can be found in the same way as in 
Section \ref{sec.polar}.

We define for all values of $i$ the polarization vectors
$\hat{U}_{(s)i}^u(k)$,  $\hat{V}_{(s)i}^u(k)$,  $\hat{U}_{(s)i}^v(k)$,  $\hat{V}_{(s)i}^v(k)$ in terms of 
$\hat{u}_{(s)i}(k)$, $\hat{v}_{(s)i}(k)$ in the following way:
\begin{eqnarray}
\label{eq.RSSpol1}
\hat{U}_{(s)i}^u(k) & = & \frac{1}{\sqrt{2}}(\hat{u}_{(s)i}(k),\, -\ii\hat{u}_{(s)i}(k)) \\
\hat{V}_{(s)i}^u(k) & = & \frac{1}{\sqrt{2}}(\hat{u}_{(s)i}(k),\, \phantom{-}\ii\hat{u}_{(s)i}(k)) \\
\hat{U}_{(s)i}^v(k) & = & \frac{1}{\sqrt{2}}(\hat{v}_{(s)i}(k),\, -\ii\hat{v}_{(s)i}(k)) \\
\label{eq.RSSpol4}
\hat{V}_{(s)i}^v(k) & = & \frac{1}{\sqrt{2}}(\hat{v}_{(s)i}(k),\, \phantom{-}\ii\hat{v}_{(s)i}(k))\ .
\end{eqnarray}

The non-zero scalar products of these polarization vectors are 
\begin{eqnarray}
\braketv{\hat{U}_{(s)i}^u(k)}{\hat{U}_{(s')j}^u(k)} \ = \ \phantom{-}\delta_{ss'}\delta_{ij}\ ,&& \qquad 
\braketv{\hat{V}_{(s)i}^u(k)}{\hat{V}_{(s')j}^u(k)} \ = \ -\delta_{ss'}\delta_{ij}, \\
\braketv{\hat{U}_{(s)i}^v(k)}{\hat{U}_{(s')j}^v(k)} \ = \ -\delta_{ss'}\delta_{ij}\ ,&& \qquad 
\braketv{\hat{V}_{(s)i}^v(k)}{\hat{V}_{(s')j}^v(k)} \ = \ \phantom{-}\delta_{ss'}\delta_{ij}.
\end{eqnarray}

The corresponding dual polarization vectors are
\begin{eqnarray}
\label{eq.RSSdpol1}
U_{(s)i}^u(k) \ =\  \frac{1}{\sqrt{2}} \left( \begin{array}{c}
u_{(s)i}(k) \\
\ii u_{(s)i}(k) \\
\end{array}\right)\ ,&&
\qquad
V_{(s)i}^u(k) \ =\  \frac{1}{\sqrt{2}} \left( \begin{array}{c}
u_{(s)i}(k) \\
-\ii u_{(s)i}(k) \\
\end{array}\right)\ , \\
\label{eq.RSSdpol2}
U_{(s)i}^v(k) \ =\  \frac{1}{\sqrt{2}} \left( \begin{array}{c}
v_{(s)i}(k) \\
\ii v_{(s)i}(k) \\
\end{array}\right)\ ,&&
\qquad
V_{(s)i}^v(k) \ =\  \frac{1}{\sqrt{2}} \left( \begin{array}{c}
v_{(s)i}(k) \\
-\ii v_{(s)i}(k) \\
\end{array}\right) \ .
\end{eqnarray}

The creation operators corresponding to the above polarization vectors can be defined as 
\begin{eqnarray}
c_{(s)i}^{U\dagger}(k) & = & \hat{U}^u_{(s)i}(k)^\alpha a_\alpha^\dagger(k),\\
c_{(s)i}^{V\dagger}(k) & = & \hat{V}^u_{(s)i}(k)^\alpha a_\alpha^\dagger(k),\\
f_{(s)i}^{U\dagger}(k) & = & \hat{U}^v_{(s)i}(k)^\alpha a_\alpha^\dagger(k),\\
f_{(s)i}^{V\dagger}(k) & = & \hat{V}^v_{(s)i}(k)^\alpha a_\alpha^\dagger(k),\qquad i=1,\dots, d_s \ .
\end{eqnarray}
These operators satisfy the (anti)commutation relations
\begin{eqnarray}
&& [ c^U_{(s)i}(k),c_{(s')j}^{U\dagger} (k') ]_\pm = 
\phantom{-}\delta_{ss'}\delta_{ij}\delta^3(k-k')\omega(k)\\
&& [ c^V_{(s)i}(k),c_{(s')j}^{V\dagger} (k') ]_\pm =  
-\delta_{ss'}\delta_{ij}\delta^3(k-k')\omega(k)\\
&& [ f^U_{(s)i}(k),f_{(s')j}^{U\dagger} (k') ]_\pm =  
-\delta_{ss'}\delta_{ij}\delta^3(k-k')\omega(k)\\
&& [ f^V_{(s)i}(k),f_{(s')j}^{V\dagger} (k') ]_\pm =  
\phantom{-}\delta_{ss'}\delta_{ij}\delta^3(k-k')\omega(k);
\end{eqnarray}
all other (anti)commutators of them are zero.
This shows that the one-particle states created by 
$c_{(s)i}^{U\dagger}(k)$ have positive scalar product with themselves,
the one-particle states created by 
$c_{(s)i}^{V\dagger}(k)$ have negative scalar product with themselves,
whereas 
$f_{(s)i}^{U\dagger}(k)$ create one-particle states 
that have negative scalar product with themselves,
and $f_{(s)i}^{V\dagger}(k)$ create one-particle states 
that have positive scalar product with themselves.

Using completeness relations analogous to (\ref{eq.c1}) and (\ref{eq.c2}), $\Phi$ can be expressed in terms of 
the above creation operators in the case of $D=D^{(n,m)}$
as
\begin{eqnarray}
&& \Phi_\alpha (x,t)=\int \frac{\intd^3 k}{\sqrt{2}(\sqrt{2\pi})^3 \omega(k)}\ \times \nonumber \\
&& \hspace{1cm}
 \Big[ e^{\ii kx}e^{\ii \omega(k)t}
    \sum_s  \sum_{i=1}^{d_s}  
\big\{ U^u_{(s)i\alpha}(k) c_{(s)i}^{U\dagger}(k) + 
V^u_{(s)i\alpha}(k) c_{(s)i}^{V\dagger}(k) \nonumber \\ 
&& \hspace{3cm} + U^v_{(s)i\alpha}(k) f_{(s)i}^{U\dagger}(k) +
V^v_{(s)i\alpha}(k) f_{(s)i}^{V\dagger}(k) \big\} \nonumber \\
&& \hspace{1cm} +e^{-\ii kx}e^{-\ii\omega(k)t}
    \sum_s \sum_{i=1}^{d_s} 
\big\{ U^u_{(s)i\alpha}(k)^* c^U_{(s)i}(k) +
V^u_{(s)i\alpha}(k)^* c^V_{(s)i}(k)  \nonumber \\
&& \hspace{3cm}  + U^v_{(s)i\alpha}(k)^* f^U_{(s)i}(k) +
V^v_{(s)i\alpha}(k)^* f^V_{(s)i}(k)
 \big\} 
\Big] \ .
\label{eq.RSSmodeexp}
\end{eqnarray}

In the case of $D=\tilde{D}^{(n)}$, 
\beq
\Phi=\phi + \phi^-,
\eeq 
where  
\begin{eqnarray}
&& \phi_\alpha (x,t)=\int \frac{\intd^3 k}{\sqrt{2}(\sqrt{2\pi})^3 \omega(k)}\ \times \nonumber \\
&& \hspace{0.5cm} \Big[e^{\ii kx}e^{\ii \omega(k)t}
    \sum_{s\in S_+}  \sum_{i=1}^{d_s}  
\big\{ U^u_{(s)i\alpha}(k) c_{(s)i}^{U\dagger}(k) 
+ V^u_{(s)i\alpha}(k) c_{(s)i}^{V\dagger}(k) \big\}  \nonumber \\
&& \hspace{0.6cm} +e^{-\ii kx}e^{-\ii\omega(k)t}
    \sum_{s\in S_+} \sum_{i=1}^{d_s} 
\big\{ U^u_{(s)i\alpha}(k)^* c^U_{(s)i}(k)   
+ V^u_{(s)i\alpha}(k)^* c^V_{(s)i}(k) \big\} 
\Big]
\label{eq.RSSmodeexp2}
\end{eqnarray}
and
\begin{eqnarray}
&& (\phi^-)_\alpha (x,t)=\int \frac{\intd^3 k}{\sqrt{2}(\sqrt{2\pi})^3 \omega(k)} \ \times \nonumber \\
&& \hspace{0.5cm} \Big[ 
e^{\ii kx}e^{\ii \omega(k)t}
\sum_{s\in S_-}  \sum_{i=1}^{d_s} 
\big\{ U^v_{(s)i\alpha}(k) f_{(s)i}^{U\dagger}(k)  
+ V^v_{(s)i\alpha}(k) f_{(s)i}^{V\dagger}(k) \big\}  \nonumber \\
&& \hspace{0.6cm}
+e^{-\ii kx}e^{-\ii\omega(k)t}
\sum_{s\in S_-} \sum_{i=1}^{d_s} 
\big\{ U^v_{(s)i\alpha}(k)^* f^U_{(s)i}(k) 
+ V^v_{(s)i\alpha}(k)^* f^V_{(s)i}(k) \big\}
\Big].
\label{eq.RSSmodeexp3}
\end{eqnarray}

The mode expansion formula (\ref{eq.RSSmodeexp}), 
together with the properties of the polarization vectors described above, shows that 
in the case of $D=D^{(n,m)}$ $\Phi$ describes $4$ kinds of particles of mass $\mass$ for any value of $s$, 
of which $2$ are physical
and $2$ are non-physical (in the sense that the one-particle states have
positive and negative scalar product with themselves, respectively).
Formulas (\ref{eq.RSSmodeexp2}) and (\ref{eq.RSSmodeexp3}) 
show that in the case of $D=\tilde{D}^{(n)}$ $\Phi$ describes
$2$ kinds of particles of mass $\mass$ for any value of $s\in S_+$ or $s\in S_-$ of which one 
is physical and one is non-physical.
The definition of the polarization vectors 
$\hat{U}_{(s)i}^u(k)$,  $\hat{V}_{(s)i}^u(k)$,  $\hat{U}_{(s)i}^v(k)$,  $\hat{V}_{(s)i}^v(k)$ 
shows that to each normal (non-RSS) one-particle state there corresponds two RSS one-particle states,
of which one is physical and one is non-physical (regarding the positivity properties of the scalar products of these states with themselves). The members of these pairs of RSS particles are distinguished in notation by the letters $U$ and $V$.  

Since the RSS auxiliary field $\Phi$ fits in the framework of Section \ref{sec.mkg}, the energy of all states 
that can be created by $\Phi$ and $\Phi^\dagger$ is positive, i.e.\ in this respect the RSS fields are not non-physical.

\subsection{Elementary higher spin fields with reversed spin-statistics relation}
\label{sec.hsgf}

In this section we turn to the definition of elementary higher spin RSS fields, which have definite spin.
For the definition of these fields we take the projection operators
\beq
\label{eq.RSSproj}
\tilde{\mc{D}}_{(s)}[\partial]\ = \ \left( \begin{array}{cc}
\mc{D}_{(s)}[\partial] & 0 \\
0 & \mc{D}_{(s)}[\partial] \\
\end{array}\right) \ ,
\eeq
where $\mc{D}_{(s)}[\partial]$ is the differential operator that appears in the definition of the normal
elementary spin $s$ field to which the RSS field will correspond. 
The elementary RSS field of spin $s$ is thus given by  
\beq
\phi_{(s)}\ = \ \tilde{\mc{D}}_{(s)}[\partial]\Phi\ .
\eeq
The block diagonal form of $\tilde{\mc{D}}_{(s)}[\partial]$ implies that 
\begin{eqnarray}
\phi_{(s)A} & = & \mc{D}_{(s)}[\partial]\Phi_A\\
\phi_{(s)B} & = & \mc{D}_{(s)}[\partial]\Phi_B\ .
\end{eqnarray}
We also define the projection operator $\tilde{\mc{D}}^-[\partial]$ in the same manner.

$\tilde{\mc{D}}_{(s)}[\partial]$ has the following action on the dual polarization vectors introduced
in Section \ref{sec.RSSpolar}:
\begin{eqnarray}
{\tilde{\mc{D}}_{(s)}[\ii k]_\alpha}^\beta  U^u_{(s')i\beta}(k) & = & \delta_{ss'} U^u_{(s)i\alpha}(k) \\
{\tilde{\mc{D}}_{(s)}[\ii k]_\alpha}^\beta  V^u_{(s')i\beta}(k) & = & \delta_{ss'} V^u_{(s)i\alpha}(k) \\
{\tilde{\mc{D}}_{(s)}[\ii k]_\alpha}^\beta U^v_{(s')i\beta}(k) & = & 0 \\
{\tilde{\mc{D}}_{(s)}[\ii k]_\alpha}^\beta V^v_{(s')i\beta}(k) & = & 0 \\
{\tilde{\mc{D}}_{(s)}[-\ii k]_\alpha}^\beta  U^u_{(s')i\beta}(k)^* & = & \delta_{ss'} U^u_{(s)i\alpha}(k)^* \\
{\tilde{\mc{D}}_{(s)}[-\ii k]_\alpha}^\beta  V^u_{(s')i\beta}(k)^* & = & \delta_{ss'} V^u_{(s)i\alpha}(k)^* \\
{\tilde{\mc{D}}_{(s)}[-\ii k]_\alpha}^\beta U^v_{(s')i\beta}(k)^* & = & 0 \\
{\tilde{\mc{D}}_{(s)}[-\ii k]_\alpha}^\beta V^v_{(s')i\beta}(k)^* & = & 0 \ .
\end{eqnarray}
Here the second four equations are obtained from the first four equations by complex conjugation.

${\tilde{\mc{D}}_{(s)}[\ii k]_\alpha}^\beta$ and ${\tilde{\mc{D}}^-[\ii k]_\alpha}^\beta$ can be expressed in terms of the polarization vectors as 
\begin{eqnarray}
\label{eq.RSSp1}
\sum_{i=1}^{d_s} \big( U^u_{(s)i\alpha}(k) \hat{U}_{(s)i}^u (k)^\beta +  V^u_{(s)i\alpha}(k) \hat{V}_{(s)i}^u (k)^\beta \big) & = & 
{ \tilde{\mc{D}}_{(s)}[\ii k]_\alpha}^\beta \\
\label{eq.RSSp2}
\sum_s \sum_{i=1}^{d_s} \big( U^v_{(s)i\alpha}(k) \hat{U}_{(s)i}^v (k)^\beta + V^v_{(s)i\alpha}(k) \hat{V}_{(s)i}^v (k)^\beta \big)
& = &  { \tilde{\mc{D}}^-[\ii k]_\alpha}^\beta \ .
\end{eqnarray}
Complex conjugation of these formulas gives
\begin{eqnarray}
\label{eq.RSSp3}
\sum_{i=1}^{d_s} \big( U^u_{(s)i\alpha}(k)^* \hat{U}_{(s)i}^u (k)^{\beta *} +  V^u_{(s)i\alpha}(k)^* \hat{V}_{(s)i}^u (k)^{\beta *} \big) & = & { \tilde{\mc{D}}_{(s)}[-\ii k]_\alpha}^\beta \\
\label{eq.RSSp4}
\sum_s \sum_{i=1}^{d_s} \big( U^v_{(s)i\alpha}(k)^* \hat{U}_{(s)i}^v (k)^{\beta *} + V^v_{(s)i\alpha}(k)^* \hat{V}_{(s)i}^v (k)^{\beta *} \big)
& = &  { \tilde{\mc{D}}^-[-\ii k]_\alpha}^\beta  \ .
\end{eqnarray}

$\tilde{\mc{D}}_{(s)}[\partial]$ and
$\tilde{\mc{D}}^-[\partial]$ form a complete set of projectors on the space of the solutions of the Klein--Gordon equation:
\beq
\sum_s {\tilde{\mc{D}}_{(s)}[\partial]_\alpha}^\beta \Phi_\beta + 
{\tilde{\mc{D}}^-[\partial]_\alpha}^\beta\Phi_\beta\ =\ \Phi_\alpha  \ .
\eeq

$\phi_{(s)}$ has the mode expansion
\begin{eqnarray}
&& \phi_{(s)\alpha} (x,t)=\int \frac{\intd^3 k}{\sqrt{2}(\sqrt{2\pi})^3 \omega(k)}\ \times \nonumber \\
&&  \hspace{1cm}
 \Big[ e^{\ii kx}e^{\ii \omega(k)t}
      \sum_{i=1}^{d_s}  
\big\{ U^u_{(s)i\alpha}(k) c_{(s)i}^{U\dagger}(k) + 
V^u_{(s)i\alpha}(k) c_{(s)i}^{V\dagger}(k) \big\} \nonumber \\ 
&& \hspace{1.1cm} +e^{-\ii kx}e^{-\ii\omega(k)t}
     \sum_{i=1}^{d_s} 
\big\{ U^u_{(s)i\alpha}(k)^* c^U_{(s)i}(k) +
V^u_{(s)i\alpha}(k)^* c^V_{(s)i}(k) \big\}
\Big];
\label{eq.RSSmex1}
\end{eqnarray}
the other modes that are present in $\Psi$ are eliminated by $\tilde{\mc{D}}_{(s)}[\partial]$.

In addition to the Klein--Gordon equation, $\phi_{(s)}$ also satisfies the differential equations
\begin{eqnarray}
\label{eq.RSSds}
\tilde{\mc{D}}_{(s)}[\partial] \phi_{(s)} & = & \phi_{(s)}   \\
\tilde{\mc{D}}_{(s')}[\partial] \phi_{(s)} & = & 0\ ,\qquad s'\ne s \\
\label{eq.d-}
\tilde{\mc{D}}^-[\partial] \phi_{(s)} & = & 0 \ .
\end{eqnarray}

The expansion (\ref{eq.RSSmex1}) shows that $\phi_{(s)}$ describes 
$2$ kinds of particles of mass $\mass$ and spin $s$, of which one is physical and one is non-physical 
(regarding the positivity properties of the scalar product).
In accordance with the statistical properties of $\Phi$ mentioned in the first part of Section \ref{sec.RSS}, 
these particles are bosonic if $s$ is half-integer and fermionic if $s$ is integer, in contrast with the usual 
spin-statistics relationship that applies to the normal fields. 

Standard complex RSS fields can be constructed in the same way as described in Sections \ref{sec.mkg} and \ref{sec.hsf}. 

A complex form of an elementary real RSS field can be constructed in the same way as 
the complex form $\Upsilon$ of a real auxiliary RSS field is constructed, i.e.\ by taking the 
combination $\upsilon_{(s)}=\frac{1}{\sqrt{2}}(\phi_{(s)A}+\ii\phi_{(s)B})$, 
and the relation between the auxiliary field $\Upsilon$ and 
$\upsilon_{(s)}$ is clearly the same as in (\ref{eq.projection}).
This field differs, of course, from the simplest kind of complex fields since its real and imaginary parts 
do not (anti)commute.

Chiral RSS fields can be defined in the same way as described at the end of Section \ref{sec.hsf}.  
In the construction of chiral fields the same projection should be applied 
on the components $\Phi_A$ and $\Phi_B$ or
$\phi_{(s)A}$ and $\phi_{(s)B}$.

The fields $\phi_{(s)}$ are elementary in the sense that they describe particles with definite mass and spin, and they
describe the minimal number, namely two, of such particles. 
It is not possible for an RSS field to describe a single particle, since that would violate 
the spin-statistics theorem.

It is also interesting to note that 
the components $\phi_{(s)A}$, $\phi_{(s)B}$ are null fields, i.e.\ 
they create only states that have zero scalar product with themselves.

\section{Scalar fields}
\label{sec.scRSS}

In this section we describe the special case of the scalar RSS field, with the aim of 
showing explicitly how the well-known Faddeev--Popov ghost field fits into the framework described in Section \ref{sec.RSS}.

\subsection{Normal scalar field}
\label{sec.nsc}

In the case of the normal (non-RSS) real scalar field  
the space of polarization vectors is one-dimensional, and we have
\beq
\epsilon=1\ ,
\eeq
\beq
u(k)=1\ ,\qquad \hat{u}(k)=1\ ,
\eeq
\beq
c^\dagger (k)=a^\dagger (k)\ .
\eeq
The projection operator is trivial ($\mc{D}[\partial]=I$), thus the auxiliary field is identical to
the physical field.
The mode expansion of the scalar field takes the form
\begin{eqnarray}
&& \phi (x,t)=
\int \frac{\intd^3 k}{\sqrt{2}(\sqrt{2\pi})^3 \omega(k)}\ 
\Big[   e^{\ii kx}e^{\ii \omega(k)t}  c^\dagger(k) 
 +  e^{-\ii kx}e^{-\ii\omega(k)t}  c(k)
\Big]\ .
\label{eq.scmodex}
\end{eqnarray}

\subsection{Fermionic scalar field} 
\label{sec.RSSsc}

The $\epsilon$ tensor for the real RSS scalar field is then 
\beq
\tilde{\epsilon}=
\left( \begin{array}{cc}
0 & \ii  \\
-\ii & 0 \\
\end{array}\right)\ ,
\eeq
and the standard polarization vectors are 
\begin{eqnarray}
\hat{U}^u(k) & = & \frac{1}{\sqrt{2}}(1,\, -\ii) \\
\hat{V}^u(k) & = & \frac{1}{\sqrt{2}}(1,\, \ii)\ . 
\end{eqnarray}
In particular, $\hat{U}^u(k)$ and $\hat{V}^u(k)$ are independent of $k$.
The non-zero scalar products of these polarization vectors are 
\begin{eqnarray}
\braketv{\hat{U}^u(k)}{\hat{U}^u(k)} = 1\ ,&& \qquad 
\braketv{\hat{V}^u(k)}{\hat{V}^u(k)} = - 1\ .
\end{eqnarray}
The corresponding dual polarization vectors are
\begin{eqnarray}
U^u(k) = \frac{1}{\sqrt{2}}
\left( \begin{array}{c}
1 \\
\ii \\
\end{array}\right)\ ,&&
\qquad
V^u(k) = \frac{1}{\sqrt{2}} 
\left( \begin{array}{c}
\phantom{-}1 \\
-\ii  \\
\end{array}\right) \ .
\end{eqnarray}

The creation operators corresponding to the above polarization vectors are 
$c^{U\dagger}(k)$, $c^{V\dagger}(k)$, which have the following non-zero anticommutators:
\begin{eqnarray}
&& [c^U(k),c^{U\dagger}(k')]_+ =  \phantom{-}\delta^3(k-k')\omega(k) \\
&& [c^V(k),c^{V\dagger}(k')]_+ =  -\delta^3(k-k')\omega(k) \ .
\end{eqnarray}

The projection operator $\mc{\tilde{D}}[\partial]$ is again just the unit operator, thus 
the mode expansion of the real RSS scalar field is 
\begin{eqnarray}
&& \phi (x,t)=
\int \frac{\intd^3 k}{\sqrt{2}(\sqrt{2\pi})^3 \omega(k)}\ \times \nonumber \\
&& \hspace{2.5cm} \Big[e^{\ii kx}e^{\ii \omega(k)t}
\{ U^u c^{U\dagger}(k) 
+ V^u c^{V\dagger}(k) \}  \nonumber \\
&& \hspace{3.1cm} +e^{-\ii kx}e^{-\ii\omega(k)t} 
\{ U^{u*} c^U(k)   
+ V^{u*} c^V(k) \} 
\Big]\ .
\label{eq.RSSsc}
\end{eqnarray}
In particular,
\begin{eqnarray}
&& \phi_A (x,t)=
\int \frac{\intd^3 k}{2(\sqrt{2\pi})^3 \omega(k)}\ \times \nonumber \\
&& \hspace{2.5cm} \Big[e^{\ii kx}e^{\ii \omega(k)t}
\{ c^{U\dagger}(k) 
+  c^{V\dagger}(k) \}  \nonumber \\
&& \hspace{3.1cm} +e^{-\ii kx}e^{-\ii\omega(k)t} 
\{  c^U(k)   
+  c^V(k) \} 
\Big]\ .
\label{eq.RSSscA}
\end{eqnarray}
and 
\begin{eqnarray}
&& \phi_B (x,t)=
\int \frac{\intd^3 k}{2(\sqrt{2\pi})^3 \omega(k)}\ \times \nonumber \\
&& \hspace{2.5cm} \Big[e^{\ii kx}e^{\ii \omega(k)t}
\{ \ii c^{U\dagger}(k) 
 -\ii c^{V\dagger}(k) \}  \nonumber \\
&& \hspace{3.1cm} +e^{-\ii kx}e^{-\ii\omega(k)t} 
\{ -\ii c^U(k)   
+  \ii c^V(k) \} 
\Big]\ .
\label{eq.RSSscB}
\end{eqnarray}

The Lagrangian (\ref{eq.LPhi}) for the real RSS scalar field is 
\beq
L=\int \intd^3 x\ 
\ii \big[ (\partial_\mu\phi_A) (\partial^\mu\phi_B)
 -\mass^2 \phi_A\phi_B \big]\ .
\eeq
By introducing the fields $c$ and $\bar{c}$ as 
\beq
\label{eq.ccbar}
\phi_A=\bar{c}\ ,\qquad \phi_B=-\ii c\ ,
\eeq
we get the usual form of the scalar ghost field Lagrangian (see e.g.\ Ref. \cite{PeskinSchroeder}) 
\beq
\label{eq.cclagr}
L=\int \intd^3 x\ 
\big[ (\partial_\mu\bar{c}) (\partial^\mu c)
 -\mass^2 \bar{c}c \big]\ .
\eeq
In (\ref{eq.ccbar}) $c$ is chosen to be imaginary and $\bar{c}$ to be real
for conformity with
the BRST transformation rules in Section 16.4 of Ref. \cite{PeskinSchroeder}.
Using (\ref{eq.feynmanprop}) 
the propagator $\brakettt{0}{\mathrm{T}c(x,t_x)\bar{c}(y,t_y)}{0}$
is found to be
\beq
\brakettt{0}{\mathrm{T}c(x,t_x)\bar{c}(y,t_y)}{0} = \int \frac{\intd^3 k\, \intd k_0}{(2\pi)^4}
\frac{\ii}{k_0^2-k^2-\mass^2+\ii\epsilon}e^{-\ii k(x-y)}e^{-\ii k_0 (t_x-t_y)} \ ,
\eeq
which, after setting $\mass=0$, is also in agreement with the standard expression (see e.g.\ Ref. \cite{PeskinSchroeder}).
For $\brakettt{0}{\mathrm{T}\bar{c}(x,t_x)c(y,t_y)}{0}$ it is found that 
\beq
\brakettt{0}{\mathrm{T}\bar{c}(x,t_x)c(y,t_y)}{0} = - \brakettt{0}{\mathrm{T}c(x,t_x)\bar{c}(y,t_y)}{0}\ .
\eeq

The complex form $\upsilon=\Upsilon=\frac{1}{\sqrt{2}}(\phi_A+\ii\phi_B)$ 
of $\phi$ is a fermionic
single-component complex Klein-Gordon 
field, with the Lagrangian 
$\int \intd^3 x\ 
\big[ (\partial_\mu\upsilon^\dagger) (\partial^\mu \upsilon)
 -\mass^2 \upsilon^\dagger \upsilon \big]$.

\section{Dirac fields}
\label{sec.dirac}

In this section we discuss the case of the Dirac field in more detail.
In particular, we also present the direct Lagrangian formulation (in which the auxiliary field is not used)
both for the normal and for the RSS Dirac field.

We use the overbar notation
\beq
\label{eq.diracconj}
\bar{\psi}^\beta=\psi_\alpha^\dagger\epsilon^{\alpha\beta}\ .
\eeq
In the following it will be seen that this coincides with the usual Dirac conjugate.

\subsection{Normal Dirac field}
\label{sec.ndirac}

Although the standard Dirac field is a complex field, it can be constructed from 
two real anticommuting Dirac fields in the simple way described at the end of Section \ref{sec.hsf}.

The projection operator for the normal Dirac field is  
\beq
\label{eq.dpr}
\mc{D}_{(1/2)}[\partial]=\frac{1}{2\mass}(\mass+\ii \gamma^\mu\partial_\mu)\ ,
\eeq 
and
\beq
\mc{D}^-[\partial]=\frac{1}{2\mass}(\mass-\ii \gamma^\mu\partial_\mu)\ .
\eeq
The real Dirac field $\phi$ thus satisfies the differential equation
\beq
2\mass (\phi-\mc{D}_{(1/2)}[\partial]\phi)=(\mass-\ii \gamma^\mu\partial_\mu)\phi=0\ ,
\eeq
which is the well-known Dirac equation (although for a real field). $\phi$ satisfies equation (\ref{eq.dmp}), i.e.\
$\mc{D}^-[\partial]\phi=0$, as well, but this is also identical to the Dirac equation.  
The construction of the complex Dirac field $\psi$ implies that it also satisfies the Dirac equation.   

In order to have the usual normalization for $\psi$, we define the real Dirac field as
$\phi=\sqrt{2\mass}\mc{D}_{(1/2)}[\partial]\Phi$ (i.e.\ the factor $\sqrt{2\mass}$ is inserted into (\ref{eq.projection})).
The equal time anticommutation relations that follow from this definition for $\phi$ and $\psi$ are 
\begin{eqnarray}
\{\phi_\alpha(x,t),\phi_\beta(x',t)\} & = & {(\gamma^0)_\alpha}^\rho \epsilon_{\rho\beta}\delta^3(x-x') 
\label{eq.dcommreal}
\end{eqnarray}
and
\begin{eqnarray}
\{\psi_\alpha(x,t),\psi_\beta(x',t)\} & = & 0 \\
\{\psi_\alpha^\dagger(x,t),\psi_\beta^\dagger(x',t)\} & = & 0 \\
\{\psi_\alpha(x,t),\psi_\beta^\dagger(x',t)\} & = & {(\gamma^0)_\alpha}^\rho \epsilon_{\rho\beta}\delta^3(x-x')\ . 
\label{eq.dcomm}
\end{eqnarray}

In Ref. \cite{toth} we took a real basis in $D^{(1,0)}$ with respect to which the gamma matrices have the form
$$
\gamma^0=\left( \begin{array}{cccc}
0 & \ii & 0 & 0\\
-\ii & 0 & 0 & 0\\
0 & 0 & 0 & \ii\\
0 & 0 & -\ii & 0\\
\end{array}\right)\quad 
\gamma^1=\left( \begin{array}{cccc}
0 & 0 & 0 & -\ii\\
0 & 0 & -\ii & 0\\
0 & -\ii & 0 & 0\\
-\ii & 0 & 0 & 0\\
\end{array}\right)\quad
\gamma^2=\left( \begin{array}{cccc}
\ii & 0 & 0 & 0\\
0 & -\ii & 0 & 0\\
0 & 0 & \ii & 0\\
0 & 0 & 0 & -\ii\\
\end{array}\right) 
$$
$$
\gamma^3=\left( \begin{array}{cccc}
0 & -\ii & 0 & 0\\
-\ii & 0 & 0 & 0\\
0 & 0 & 0 & \ii\\
0 & 0 & \ii & 0\\
\end{array}\right)\ , 
$$
and 
$\epsilon$ has the canonical form
$$
\epsilon=
\left( \begin{array}{cccc}
0 & \ii & 0 & 0 \\
-\ii & 0 & 0 & 0\\
0 & 0 & 0 & \ii\\
0 & 0 & -\ii & 0\\
\end{array}\right) .
$$
This shows, in particular, that the gamma matrices are purely imaginary with respect to the invariant complex conjugation, and thus the 
differential operators $\mc{D}_{(1/2)}[\partial]$ and $\mc{D}^-[\partial]$ introduced above are real.
It is also worth noting that 
${(\gamma^\mu)_\alpha}^\rho \epsilon_{\rho\beta}$ and ${(\gamma^\mu)_\rho}^\alpha \epsilon^{\rho\beta}$  are symmetric in $\alpha$ and $\beta$. 

The standard dual polarization vectors at $k=0$ in Ref. \cite{toth} were
\beq
u_1=\frac{1}{2}\left( \begin{array}{c}
-1  \\
-\ii \\
-1 \\
-\ii
\end{array}\right),\qquad
u_2=\frac{1}{2}\left( \begin{array}{c}
1  \\
\ii \\
-1 \\
-\ii
\end{array}\right),\qquad
v_1=\frac{1}{2}\left( \begin{array}{c}
1  \\
-\ii \\
-1 \\
\ii
\end{array}\right),\qquad
v_2=\frac{1}{2}\left( \begin{array}{c}
1  \\
-\ii \\
1 \\
-\ii
\end{array}\right).
\eeq
The corresponding polarization vectors are 
\begin{eqnarray}
\hat{u}_1 & = & \frac{1}{2}(-1,\ii,-1,\ii) \\
\hat{u}_2 & = & \frac{1}{2}(1,-\ii,-1,\ii) \\
\hat{v}_1 & = & \frac{1}{2}(1,\ii,-1,-\ii) \\
\hat{v}_2 & = & \frac{1}{2}(1,\ii,1,\ii) \ .
\end{eqnarray}

The Dirac equation can be derived from the Lagrangian 
\beq
\label{eq.diraclagr}
L= \int \intd^3 x\ \mc{L}  =  -\int \intd^3 x\ \bar{\psi}(\mass-\ii\gamma^\mu\partial_\mu)\psi\ ,  
\eeq
where $\psi$ is understood to be an anticommuting field. 
The canonical momentum field corresponding to $\psi_\alpha$ is  
\beq
\label{eq.dpi}
\tilde{\pi}^\alpha= \frac{\partial\mc{L}}{\partial(\partial_t\psi_\alpha)} = -\ii\psi_\gamma^\dagger{(\gamma^0)_\beta}^\alpha \epsilon^{\gamma\beta} = -\ii (\bar{\psi}\gamma^0)^\alpha  \ .
\eeq
The canonical anticommutation relation
\beq
\{ \psi_\alpha(x,t), \tilde{\pi}^\beta (x',t) \} = -\ii{\delta_\alpha}^\beta \delta^3(x-x') 
\eeq
together with (\ref{eq.dpi}),
is equivalent to (\ref{eq.dcomm}).
The Hamiltonian that can be derived from (\ref{eq.diraclagr}) is
\beq
H=\int \intd^3 x\ \bar{\psi} (\mass-\ii\gamma^x\partial_x) \psi\ .  
\eeq

In the representation described above, the matrix of $\epsilon$ is the same as the matrix of $\gamma^0$, 
thus $\bar{\psi}$ is just the Dirac conjugate
of $\psi$, and the Lagrangian (\ref{eq.diraclagr}) is also the usual Lagrangian for the Dirac field.   
In addition, ${(\gamma^0)_\alpha}^\rho \epsilon_{\rho\beta}=\delta_{\alpha\beta}$ holds, thus the anticommutation relation 
(\ref{eq.dcomm}) agrees with the usual one.
The basis in which $\gamma^\mu$ takes the above form is related to the Weyl basis, in which $\gamma^\mu$ has the form
\beq
\gamma^0=\left( \begin{array}{cc}
0 & I \\
I & 0 \\
\end{array}\right),\qquad
\gamma^i=\left( \begin{array}{cc}
0 & \sigma_i \\
-\sigma_i & 0 \\
\end{array}\right),\qquad i=1,2,3\ ,
\eeq
where where $I$ denotes the $2\times 2$ identity matrix,
by a unitary (but not real) matrix, therefore the transition between these two bases leaves the form
$\{\psi_\alpha(x,t),\psi_\beta^\dagger(x',t)\}  =  \delta_{\alpha\beta}\delta^3(x-x')$ of (\ref{eq.dcomm}) invariant.
The usual form $\bar{\psi}=\psi^\dagger \gamma^0$ of the Dirac conjugate also remains unchanged, since this form 
of the Dirac conjugate is unchanged under any basis change that is described by a unitary matrix. 
From these observations we can conclude that the Dirac field in our formulation is equivalent to the usual formulation in the Weyl representation.

We note that the formula (\ref{eq.diracconj}) appears more suitable as a definition of the Dirac conjugate than  
the usual
$\bar{\psi}=\psi^\dagger \gamma^0$, since the latter is not obviously Lorentz-covariant, in contrast with
(\ref{eq.diracconj}). In addition, (\ref{eq.diracconj}) is meaningful for all the fields studied in this paper, 
not only for the Dirac field, and thus  
it can also be considered as a wide generalization of the Dirac conjugation. 
Of course, it should be kept in mind that in (\ref{eq.diracconj}) 
it is assumed that the indices label vector and tensor components with respect to a real basis in $D$.
If a complex basis (like for example the Weyl basis in $D^{(1,0)}$) is used, then the formula for the Dirac conjugation takes the form
$\bar{\psi}^\beta={\mc{J}_\alpha}^\gamma   \psi_\gamma^\dagger  \epsilon^{\alpha\beta}$, where
$\mc{J}$ is the matrix of the invariant complex conjugation in this basis.
$\mc{J}$ is defined by the formula $(v^*)_\alpha = {\mc{J}_\alpha}^\beta v_\beta^*$, where 
the ${}^*$ on the left hand side denotes the invariant complex conjugation in $D$, whereas on the right hand side
${}^*$ is the componentwise complex conjugation. Thus $\mc{J}$ is the matrix that describes the relation between 
the invariant complex conjugation and the componentwise complex conjugation in the actually used basis.

\subsection{Bosonic Dirac field}
\label{sec.RSSdirac}

Taking into consideration (\ref{eq.dpr}) and the general definition (\ref{eq.RSSproj}) 
of projection operators for the RSS fields, we have
\beq
\tilde{\mc{D}}_{(1/2)}[\partial] = 
\frac{1}{2\mass}(\mass+\ii \tilde{\gamma}^\mu\partial_\mu)\ ,  
\eeq
where
$\tilde{\gamma}^\mu$ is 
\beq
\tilde{\gamma}^\mu = 
\left( \begin{array}{cc}
\gamma^\mu & 0 \\
0 & \gamma^\mu \\
\end{array}\right)\ .
\eeq
The real
RSS Dirac field $\phi=\sqrt{2\mass}\tilde{\mc{D}}_{(1/2)}[\partial]\Phi$ satisfies the differential equation 
(\ref{eq.RSSds}), i.e.\
\beq
\label{eq.RSSdirac}
2\mass (\phi-\tilde{\mc{D}}_{(1/2)}[\partial]\phi) =  (\mass-\ii \tilde{\gamma}^\mu\partial_\mu) \phi = 0\ ,
\eeq
which is 
equivalent to the Dirac equation for the two components $\phi_A$, $\phi_B$:
\begin{eqnarray}
\label{eq.RSSdA}
(\mass-\ii \gamma^\mu\partial_\mu) \phi_A & = & 0 \\
\label{eq.RSSdB}
(\mass-\ii \gamma^\mu\partial_\mu) \phi_B & = & 0\ .
\end{eqnarray}
Equation (\ref{eq.d-}) takes the form 
$(2\mass)\tilde{\mc{D}}^-[\partial]\phi =  (\mass-\ii \tilde{\gamma}^\mu\partial_\mu) \phi = 0$, 
thus is identical to (\ref{eq.RSSdirac}).
The complex RSS Dirac field $\psi$ and its components $\psi_A$, $\psi_B$  also satisfy equations
(\ref{eq.RSSdirac}) and (\ref{eq.RSSdA}), (\ref{eq.RSSdB}).

From the commutation relations (\ref{eq.ac1}) of the auxiliary field $\Phi$ and from the relation 
$\phi=\sqrt{2\mu}\tilde{\mc{D}}_{(1/2)}[\partial]\Phi$ it follows that
$\phi$ and $\psi$ satisfy the commutation relations 
\begin{eqnarray}
\label{eq.dRSScommreal}
[\phi_\alpha(x,t) , \phi_\beta (x',t) ] & = & - {(\tilde{\gamma}^0)_\beta}^\rho \tilde{\epsilon}_{\rho\alpha}\delta^3(x-x') 
\end{eqnarray} 
and
\begin{eqnarray}
[\psi_\alpha(x,t) , \psi_\beta (x',t) ] & = & 0 \\
{}[\psi^\dagger_\alpha(x,t) , \psi^\dagger_\beta (x',t) ] & = & 0 \\
\label{eq.dRSScomm}
[\psi_\alpha(x,t) , \psi^\dagger_\beta (x',t) ] & = & - {(\tilde{\gamma}^0)_\beta}^\rho \tilde{\epsilon}_{\rho\alpha}\delta^3(x-x')\ , 
\end{eqnarray}
which imply the commutation relations
\begin{eqnarray}
[\phi_{A\alpha}(x,t), \phi_{B\beta} (x',t) ] & = & \ii {(\gamma^0)_\beta}^\rho \epsilon_{\rho\alpha} \delta^3(x-x')
\end{eqnarray}
and
\begin{eqnarray}
[\psi_{A\alpha}(x,t), \psi_{B\beta}^\dagger (x',t) ] & = & 
\phantom{-}\ii {(\gamma^0)_\beta}^\rho \epsilon_{\rho\alpha} \delta^3(x-x')\\
{} [\psi_{B\alpha}(x,t), \psi_{A\beta}^\dagger (x',t) ] & = & 
-\ii {(\gamma^0)_\beta}^\rho \epsilon_{\rho\alpha} \delta^3(x-x')
\end{eqnarray} 
for   $\phi_A$, $\phi_B$ and $\psi_A$, $\psi_B$. 
Other similar commutators are zero.
It is worth noting that ${(\tilde{\gamma}^\mu)_\alpha}^\rho \tilde{\epsilon}_{\rho\beta}$ and
${(\tilde{\gamma}^\mu)_\rho}^\alpha \tilde{\epsilon}^{\rho\beta}$ are antisymmetric in $\alpha$ and $\beta$.

The equation of motion
$(\mass-\ii \tilde{\gamma}^\mu\partial_\mu) \psi = 0$ can be derived from the Lagrangian  
\beq
\label{eq.RSSdl}
L= -\int \intd^3 x\ \bar{\psi} (\mass-\ii \tilde{\gamma}^\mu\partial_\mu) \psi \ .
\eeq
In terms of the components $\psi_A$, $\psi_B$, this Lagrangian takes the form 
\beq
L= -\int \intd^3 x\ \big[ \ii \epsilon^{\alpha\beta} \psi_{A\alpha}^\dagger 
{(\mass-\ii \gamma^\mu\partial_\mu)_{B\beta}}^{B\gamma} \psi_{B\gamma}
-\ii \epsilon^{\alpha\beta}\psi_{B\alpha}^\dagger 
{(\mass-\ii \gamma^\mu\partial_\mu)_{A\beta}}^{A\gamma} \psi_{A\gamma}
\big] \ .
\eeq

The canonical momentum corresponding to $\psi_\alpha$ is  
\beq
\label{eq.dRSSpi}
\tilde{\pi}^\alpha=\ii\psi^\dagger_\rho {(\tilde{\gamma}^0)_\delta}^\alpha \tilde{\epsilon}^{\rho\delta}\ .
\eeq
The canonical commutation relation
\beq
[ \psi_\alpha(x,t), \tilde{\pi}^\beta (x',t) ] = \ii {\delta_\alpha}^\beta\delta^3(x-x')\ ,
\eeq
together with (\ref{eq.dRSSpi}),
reproduces (\ref{eq.dRSScomm}).

The Hamiltonian that can be obtained from (\ref{eq.RSSdl}) is 
\beq
H= \int \intd^3 x\ \bar{\psi} (\mass-\ii \tilde{\gamma}^x\partial_x) \psi \ . 
\eeq

The complex form $\upsilon$ of the real RSS Dirac field has the same Lagrangian 
as the normal complex Dirac field, i.e.\
$ -\int \intd^3 x\ \bar{\upsilon}(\mass-\ii\gamma^\mu\partial_\mu)\upsilon$.
It is not difficult to verify that the canonical commutation relations following from this Lagrangian are the same as those that follow 
from the construction of $\upsilon$ as $\upsilon= \mc{D}_{(1/2)}[\partial] \Upsilon$.

\section{Summary}
\label{sec.conclusion}

In this paper a construction of free massive quantum fields with arbitrary spin and reversed spin-statistics relation
was presented. In relation to their spin-statistics property, these fields (abbreviated as RSS fields) 
also have the characteristic property that they  
create one-particle states that form pairs consisting of one state that has positive scalar product with itself and 
another one that has negative scalar product with itself.

For the construction a framework introduced in Ref. \cite{toth} for higher spin fields was used. 
This involves the definition of 
auxiliary higher spin fields, which are constrained only by the Klein-Gordon equation and which are quantized canonically. 
An elementary higher spin field is obtained from an auxiliary field by the application of a suitable projection operator, 
which is a differential operator and is closely related to the propagator of the field. 

In the framework of Ref. \cite{toth} a higher spin field is specified if certain data, including the representation of 
$SL(2,\CC)$ according to which the field transforms, an invariant tensor $\epsilon^{\alpha\beta}$, and the projection operator applied to the auxiliary field, is given. 
In the presented construction of RSS fields these data are created 
from similar data specifying normal higher spin fields.
The details of the construction show that for any normal field there exists a corresponding RSS field with the same spin.

The $SL(2,\CC)$ representation for an RSS field, related to some normal field,
is taken to be the direct sum of two copies of the 
representation according to which the normal field transforms, thus any RSS field $\phi$ has two 
parts $\phi_{A}$ and $\phi_{B}$ corresponding to these copies. 
The statistical properties of a field depend on whether the tensor $\epsilon^{\alpha\beta}$ belonging to it
is symmetric or antisymmetric,
and the doubling of the representation according to which the field transforms makes it possible to define an $\epsilon^{\alpha\beta}$ tensor, in terms of the original $\epsilon^{\alpha\beta}$ tensor that belongs to the 
normal field, 
with symmetry property opposite to that of the original $\epsilon^{\alpha\beta}$ tensor. 
It is this definition by which the reversed spin-statistics property of the RSS fields is achieved. 
The projection operator for the RSS field is just the original projection operator acting separately on the 
two parts of the RSS field, 
hence these two parts satisfy in themselves the same field equations as the original normal field. 

As mentioned above, the RSS one-particle states form pairs; 
for each one-particle state of the normal field there is a corresponding pair of RSS one-particle states of 
which one has positive scalar product with itself and one has negative scalar product with itself.
The Feynman propagators, the (anti)commutators and the Green functions 
of the RSS field can also be expressed in terms of those of the normal field in a straightforward way.

The construction is based on real fields; with a real normal field a real RSS field is associated,
and complex fields are taken primarily to have the
form $\psi=\frac{1}{\sqrt{2}}(\phi_1 +\ii\phi_2)$, 
where $\phi_1$ and $\phi_2$ are two independent, i.e.\ (anti)commuting, copies of a real field. 
Such complex fields have a very simple relation to real fields, nevertheless although 
the two parts $\phi_A$ and $\phi_B$ of a real RSS field
do not (anti)commute, they 
can also be combined into a complex RSS field $\upsilon=\frac{1}{\sqrt{2}}(\phi_A +\ii\phi_B)$. 
These $\upsilon$ fields 
can be regarded as the fields that one obtains if one 
reverses the commutation properties of complex normal fields, 
leaving $\epsilon^{\alpha\beta}$ and the projection operator unchanged, therefore they 
are natural equivalent forms of the real RSS fields.

As special cases the scalar field and the Dirac field were discussed. 
In the case of the scalar field the aim was to show how the 
scalar Faddeev--Popov ghost field fits into the framework presented in this paper. 
The bosonic Dirac field was discussed because it is the simplest bosonic RSS field and because of the importance of
the (fermionic) Dirac field in particle physics. It was shown that the RSS Dirac field admits a first order
Lagrangian formulation (without the auxiliary field) similar to that of the normal Dirac field. 
It was also pointed out that by making use of the $\epsilon^{\alpha\beta}$ tensor the Dirac conjugation can be written in an obviously covariant form, and in this form it can be generalized naturally to any other field described in the framework of
Ref. \cite{toth} and of the present paper. 

Regarding further work, it would obviously be interesting to find applications of the RSS fields, 
different from their known application as ghost fields in the Faddeev--Popov--DeWitt and related methods.

\section*{Acknowledgments}

I would like to thank the reviewer for bringing the possibility of equivalent complex forms of real RSS fields to my attention. 
I also acknowledge support by an MTA Lend\"ulet grant.

\appendix

\section{Feynman propagators, Green functions, (anti)commutators}
\label{sec.feynman}

\renewcommand{\theequation}{A.\arabic{equation}}

The Green function of a real multi-component Klein--Gordon field is
\beq
\label{eq.green0}
\brakettt{0}{\Phi_\alpha(x,t_x)\Phi_\beta(y,t_y)}{0}=  \epsilon_{\alpha\beta} G(x-y,t_x-t_y)\ ,
\eeq
where the function $G$, defined by
\beq
G(x,t)=\int \frac{\intd^3 k}{2(2\pi)^3 \omega(k)} e^{-\ii kx}e^{-\ii \omega(k) t}\ ,
\eeq
is the Green function of the usual Klein--Gordon field.

The
Feynman propagator is
\beq
\label{eq.f01}
\brakettt{0}{\mathrm{T}\Phi_\alpha(x,t_x)\Phi_\beta(y,t_y)}{0}=  \epsilon_{\alpha\beta} D_F(x-y,t_x-t_y)\ ,
\eeq
where the function $D_F$, defined as
\beq
\label{eq.f02}
D_F(x,t)=
\int \frac{\intd^3 k\, \intd k_0}{(2\pi)^4}\, \frac{\ii }{k_0^2-k^2-\mass^2+\ii\epsilon}e^{-\ii kx}e^{-\ii k_0t}\ ,
\eeq
is the Feynman propagator of the usual Klein--Gordon field, and $T$ denotes time ordering.

The (anti)commutator $[\Phi_\alpha(x,t_x),\Phi_\beta(x,t_y)]_\pm$
is
\beq
\label{eq.comm}
[\Phi_\alpha(x,t_x),\Phi_\beta(x,t_y)]_\pm = \epsilon_{\alpha\beta}
[G(x-y,t_x-t_y)-G(y-x,t_x-t_y)]\ .
\eeq

The formulas (\ref{eq.green0}), (\ref{eq.f01}) and (\ref{eq.comm}) 
apply also to the RSS multi-component Klein--Gordon fields,  with the replacement $\epsilon \to \tilde{\epsilon}$.

The (anti)commutation relations of an elementary higher spin field can be derived from those of the auxiliary field, taking into consideration the 
relation (\ref{eq.projection}) between the elementary and the auxiliary field.
One obtains that
\begin{eqnarray}
&& [\phi_{(s)\alpha}(x,t_x),\phi_{(s)\beta}(y,t_y)]_\pm =  \nonumber\\
&& \hspace{1cm} = [\phi_{(s)\alpha}(x,t_x),\Phi_\beta(y,t_y)]_\pm \nonumber\\
&& \hspace{1cm} = [{\mc{D}_{(s)}[\partial_{(t_x,x)}]_\alpha}^\delta  \Phi_\delta(x,t_x) , \Phi_\beta(y,t_y)]_\pm \nonumber\\
&& \hspace{1cm} = {\mc{D}_{(s)}[\partial_{(t_x,x)}]_\alpha}^\delta \epsilon_{\delta\beta} [G(x-y,t_x-t_y)-G(y-x,t_y-t_x)]\ .
\label{eq.anticomm}
\end{eqnarray}
Here the subscript in $\partial_{(t_x,x)}$ indicates the variables with respect to which the differentiation should be done.

For the Green function of $\phi_{(s)}$ one obtains in a similar way that 
\begin{eqnarray}
&& \brakettt{0}{\phi_{(s)\alpha}(x,t_x) \phi_{(s)\beta}(y,t_y)}{0} \nonumber\\
&& \hspace{2cm} = \brakettt{0}{\phi_{(s)\alpha}(x,t_x)\Phi_\beta(y,t_y)}{0} \nonumber\\
&& \hspace{2cm} = \brakettt{0}{{\mc{D}_{(s)}[\partial_{(t_x,x)}]_\alpha}^\delta  \Phi_\delta(x,t_x)  \Phi_\beta(y,t_y)}{0} \nonumber\\
&& \hspace{2cm} = {\mc{D}_{(s)}[\partial_{(t_x,x)}]_\alpha}^\delta \epsilon_{\delta\beta} G(x-y,t_x-t_y)\ .
\label{eq.green}
\end{eqnarray}

The covariant part of the Feynman propagator 
\begin{eqnarray}
&& \brakettt{0}{\mathrm{T}\phi_{(s)\alpha}(x,t_x) \phi_{(s)\beta}(y,t_y)}{0} \nonumber \\
&& \hspace{2cm} = \brakettt{0}{\mathrm{T}\phi_{(s)\alpha}(x,t_x)\Phi_\beta(y,t_y)}{0} \nonumber\\
&& \hspace{2cm} = \brakettt{0}{\mathrm{T}{\mc{D}_{(s)}[\partial_{(t_x,x)}]_\alpha}^\delta  \Phi_\delta(x,t_x)  \Phi_\beta(y,t_y)}{0} 
\end{eqnarray}
of $\phi_{(s)}$ is
\beq
\label{eq.psiprop}
{\mc{D}_{(s)}[\partial_{(t_x,x)}]_\alpha}^\delta \epsilon_{\delta\beta} D_F(x-y,t_x-t_y)\ ,
\eeq
where $D_F$ denotes the Feynman propagator of the scalar Klein-Gordon field.
(\ref{eq.psiprop}) can also be written as
\beq
\label{eq.feynmanprop}
D_F(x-y,t_x-t_y)_{\alpha\beta}^C=\int \frac{\intd^3 k\, \intd k_0 }{(2\pi)^4} \frac{\ii{\mc{D}_{(s)}[-\ii (k_0,k)]_\alpha}^\delta \epsilon_{\delta\beta}}{k_0^2-k^2-\mass^2+\ii\epsilon}
e^{-\ii k(x-y)} e^{-\ii k_0(t_x-t_y)} \ .
\eeq
The notation $\mc{D}_{(s)}[-\ii (k_0,k)]$ indicates that $\partial_\mu$ is replaced
in the expression for $\mc{D}_{(s)}[\partial]$ by the momentum covector
$-\ii (k_0,k)$.

The formulas above---with $\tilde{\epsilon}$ and $\tilde{\mc{D}}_{(s)}[\partial]$, of course---apply also to 
the elementary RSS fields.

Feynman propagators, Green functions and (anti)commutation relations for complex or chiral fields can be obtained in a straightforward way 
from those of the real fields.

\section{}
\label{sec.appb}

\renewcommand{\theequation}{B.\arabic{equation}}

The existence of the vectors $u_{(s)i}$, $v_{(s)i}$
described in Section \ref{sec.polar} was shown in Appendix A of Ref. \cite{toth}, although it was not explicitly demonstrated that the constructions described there
are general in the sense that they do not imply any restriction on the choice of the invariant complex conjugation and of $\epsilon^{\alpha\beta}$ in $D^{(n,m)}$ or in $\tilde{D}^{(n)}$.  
In the following we present a derivation of this statement.

The basis vectors $e_1$, $e_2$, $e_3$, $e_4$ introduced in Appendix A of Ref. \cite{toth} for $(1/2,0)$ and $(0,1/2)$ can be redefined as $e_1 \to \alpha e_1$, $e_2\to \alpha e_2$ 
$e_3 \to \beta e_3$, $e_4\to \beta e_4$,
with any non-zero $\alpha,\beta\in \CC$, without 
changing the matrices of the generators of the $SL(2,\CC)$ algebra. 
The complex conjugation relations $e_1^*=-e_4$, $e_2^* = e_3$ will then be preserved if we  
redefine the invariant complex conjugation $J$ from $(1/2,0)$ to $(0,1/2)$ as 
$J \to \frac{\beta}{\alpha^*}J$.
The matrices of the invariant bilinear forms $\epsilon^L$ and $\epsilon^R$ on 
$(1/2,0)$ and $(0,1/2)$ are also preserved if we
redefine them as $\epsilon^L\to \frac{1}{\alpha^2}\epsilon^L$, 
$\epsilon^R\to \frac{1}{\beta^2}\epsilon^R$, and $\epsilon$ remains imaginary.
The $\eta^\mu$ and $\bar{\eta}^\mu$ tensors, from which the gamma matrices are composed, should 
be redefined as $\eta^\mu\to \frac{\alpha}{\beta}\eta^\mu$, 
$\bar{\eta}^\mu\to \frac{\beta}{\alpha}\bar{\eta}^\mu$; this leaves their matrices 
(and thus also those of $\gamma^\mu$ and $\gamma^5$) unchanged. 
The vectors $u_{(1/2)i}$ and $v_{(1/2)i}$ are defined in terms of $e_1$, $e_2$, $e_3$, $e_4$, therefore their properties---including (\ref{eq.ortho1})---also remain unchanged.

The $\epsilon$ tensors, complex conjugations, generalized gamma matrices and various 
special vectors
for $(n/2,m/2)$, $(m/2,n/2)$, $(n/2,n/2)$  and $D^{(n,m)}$ are constructed 
in Appendix A of Ref. \cite{toth}
from those of $(1/2,0)$ and $(0,1/2)$,
thus their relations (including various scalar products and matrix elements) remain unchanged
by the redefinitions above. On the other hand, the complex conjugation
$J: (n/2,m/2)\to (m/2,n/2)$ gets redefined as 
$J\to \frac{\beta^n \alpha^m}{(\alpha^*)^n(\beta^*)^m}J$, 
the $\epsilon^L$ tensor on $(n/2,m/2)$ as 
$\epsilon^L\to\frac{1}{\alpha^{2n}\beta^{2m}}\epsilon^L$, and
the $\epsilon^R$ tensor on $(m/2,n/2)$ as
$\epsilon^R\to\frac{1}{\alpha^{2m}\beta^{2n}}\epsilon^R$,
therefore by a suitable choice of $\alpha$ and $\beta$ any of the (real or imaginary) invariant 
$\epsilon$ tensors and invariant complex conjugations in $D^{(n,m)}$
or in $\tilde{D}^{(n)}$ 
can be reached 
(see also the second part of Section \ref{sec.lorentz} for the characterization of the invariant $\epsilon$ tensors and complex conjugations). 
This means that the constructions described in Appendix A of Ref. \cite{toth}
can be performed in the same form for any choice of the invariant complex conjugation and of $\epsilon$ in $D^{(n,m)}$ or in $\tilde{D}^{(n)}$.

\end{document}